\documentclass{article}

\usepackage{arxiv}
\usepackage[utf8]{inputenc}
\usepackage[T1]{fontenc}
\usepackage{hyperref}
\usepackage{url}
\usepackage{booktabs}
\usepackage{amsfonts}
\usepackage{nicefrac}
\usepackage{microtype}
\usepackage{graphicx}
\usepackage{natbib}
\usepackage{doi}
\usepackage{caption}
\usepackage{subcaption}
\usepackage{algorithm}
\usepackage{algorithmic}
\usepackage{amsmath}

\hypersetup{colorlinks=true,linkcolor=blue,citecolor=blue,urlcolor=blue}

\title{Representing and Animating Piano-Hinged Solids as Non-Manifold Cell Complexes: The Magic Photo Cube}

\author{
\href{https://orcid.org/0000-0003-3618-4166}{\includegraphics[scale=0.06]{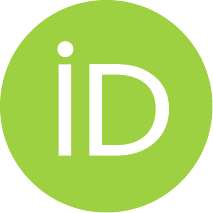}\hspace{1mm}Ergun Akleman} \\
Visual Computing \& Computational Media, SPVFA \\
Computer Science and Engineering, COE \\
Texas A\&M University, USA \\
\texttt{ergun@tamu.edu}
}

\date{November 2026}
\begin{document}\maketitle

\begin{abstract}
Piano-hinged solids are articulated structures consisting of rigid solid components connected along edges, allowing relative rotations while preserving their connectivity. We propose representing such structures as a class of edge-connected non-manifold cell complexes, in which the components and their hinge relationships are encoded through cell incidences rather than separate geometric hinge elements. This representation provides a unified framework for describing the geometry, connectivity, and admissible motions of articulated solids. We demonstrate the approach using the magic photo cube, an assembly of eight cubical cells connected by eight edge hinges. Each cell maintains a persistent identity and local coordinate frame, while successive folding operations transport hinge axes through rigid transformations. We reconstruct two distinct three-stage folding sequences and analyze their resulting configurations. Although the sequences produce compact cubes at different positions and orientations in a fixed reference frame, numerical analysis demonstrates that their terminal configurations are rigidly equivalent as labeled, oriented assemblies under a single proper rigid transformation. The example illustrates how an edge-connected non-manifold representation can support geometric modeling, animation, and configuration analysis while preserving the identities and incidence relationships of the constituent cells. An accompanying interactive web-based simulator, implemented
in JavaScript and WebGL, is publicly available at
\url{https://people.tamu.edu/~ergun/software/magic_cube}.
\end{abstract}

\begin{figure*}[t]
\centering
\begin{subfigure}[t]{0.19\textwidth}
\centering
\includegraphics[width=\linewidth]{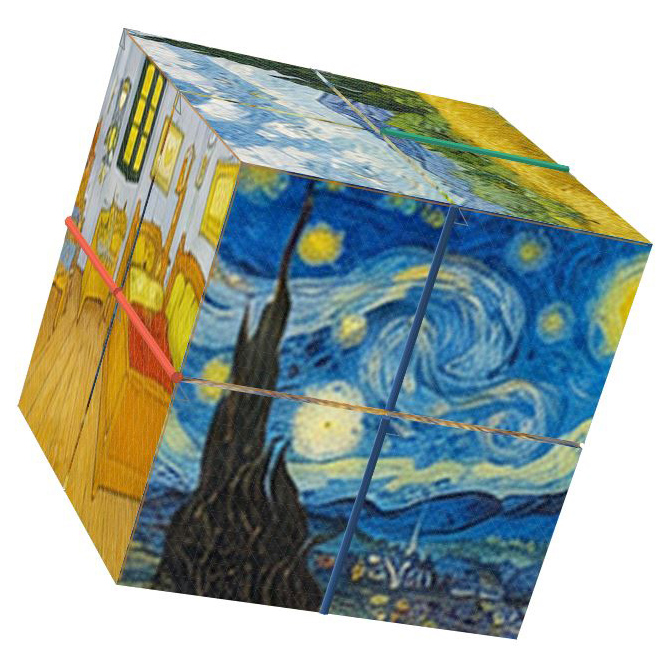}
\caption{Another View of Initial Configuration.}
\end{subfigure}
\hfill
\begin{subfigure}[t]{0.19\textwidth}
\centering
\includegraphics[width=\linewidth]{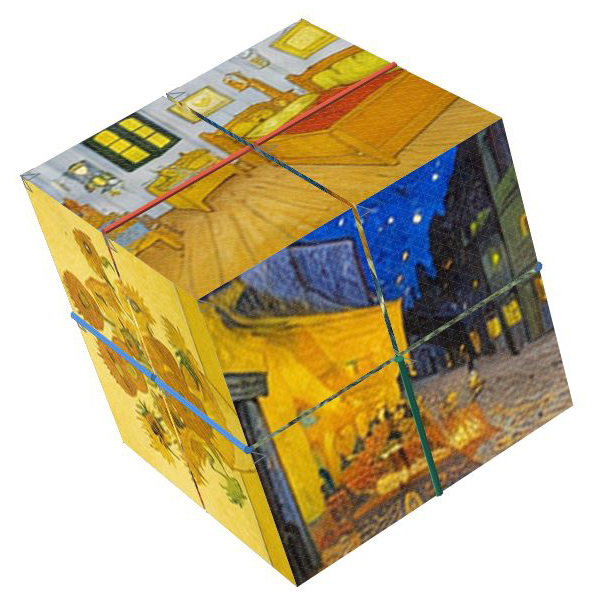}
\caption{Another View of Initial Configuration.}
\end{subfigure}
\hfill
\begin{subfigure}[t]{0.19\textwidth}
\centering
\includegraphics[width=\linewidth]{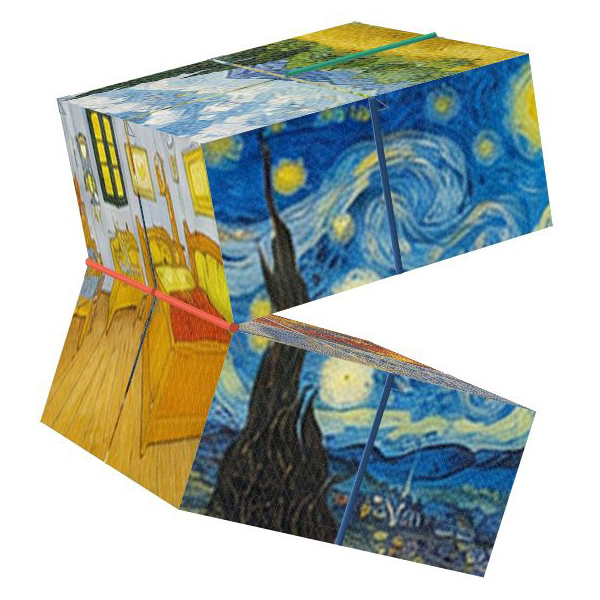}
\caption{A View from Stage 1.}
\end{subfigure}
\hfill
\begin{subfigure}[t]{0.19\textwidth}
\centering
\includegraphics[width=\linewidth]{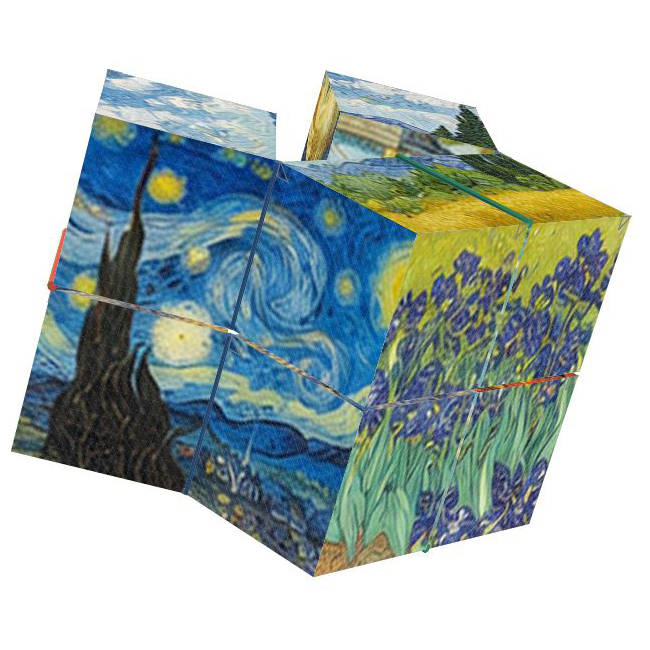}
\caption{Another View from Stage 1.}
\end{subfigure}
\hfill
\begin{subfigure}[t]{0.19\textwidth}
\centering
\includegraphics[width=\linewidth]{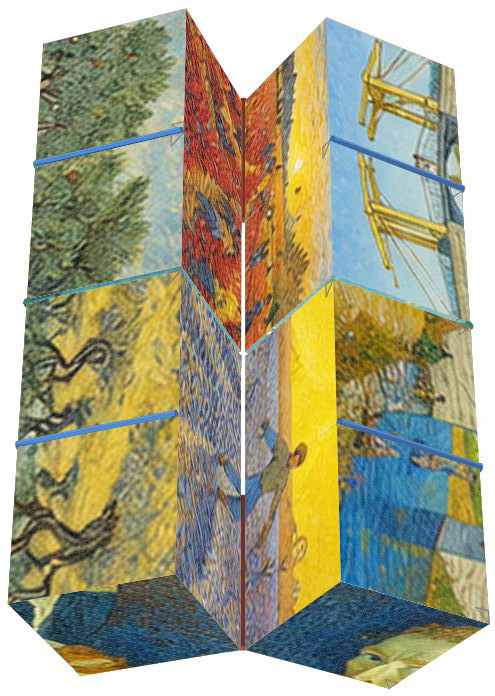}
\caption{A View from Stage 2.}
\end{subfigure}
\hfill
\begin{subfigure}[t]{0.19\textwidth}
\centering
\includegraphics[width=\linewidth]{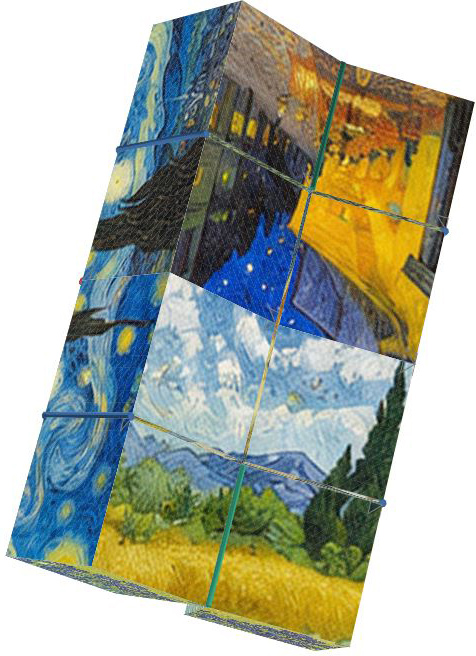}
\caption{Another View from Stage 2.}
\end{subfigure}
\hfill
\begin{subfigure}[t]{0.19\textwidth}
\centering
\includegraphics[width=\linewidth]{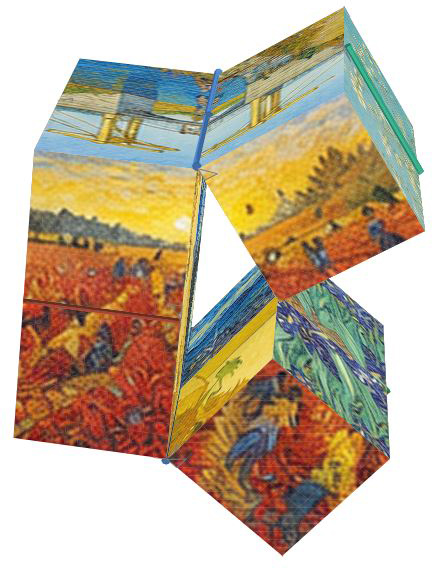}
\caption{A View from Stage 3.}
\end{subfigure}
\hfill
\begin{subfigure}[t]{0.19\textwidth}
\centering
\includegraphics[width=\linewidth]{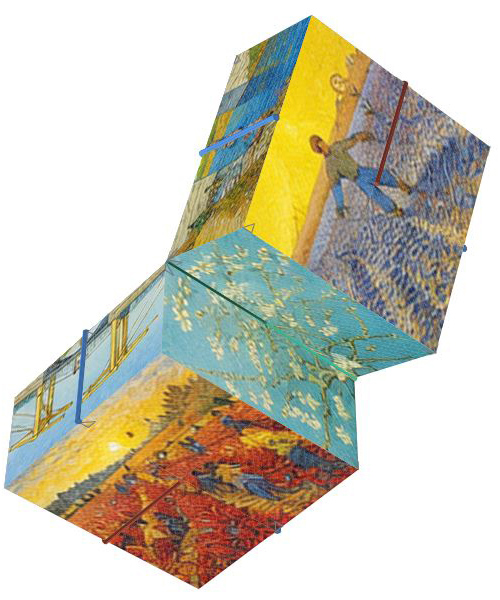}
\caption{Another View from Stage 3.}
\end{subfigure}
\hfill
\begin{subfigure}[t]{0.19\textwidth}
\centering
\includegraphics[width=\linewidth]{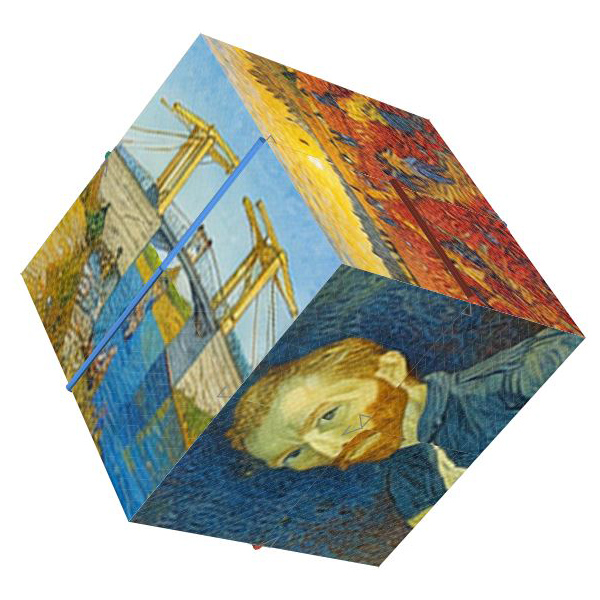}
\caption{A View of the final configuration.}
\end{subfigure}
\hfill
\begin{subfigure}[t]{0.19\textwidth}
\centering
\includegraphics[width=\linewidth]{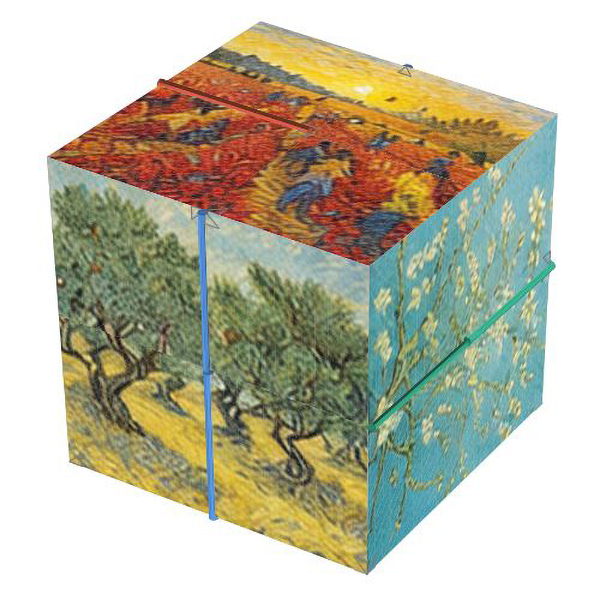}
\caption{Another View of the final configuration.}
\end{subfigure}
\hfill
\caption{Van Gogh-textured magic photo cube: ten illustrative configurations. }
\label{fig:vg_cube}
\end{figure*}

\section{Introduction and Motivation}

Articulated solids consist of rigid components connected through geometric constraints that permit relative motion while preserving their connectivity. Representing such structures requires a unified description of their geometry, topology, and allowable transformations. In this paper, we investigate piano-hinged solids as a particular class of edge-connected non-manifold cell complexes, in which hinge relationships are represented through the incidence of solid cells along selected edges. This perspective eliminates the need to model hinges as separate geometric components and provides a natural foundation for defining and analyzing articulated motions. We introduce the motivation for this representation, discuss its implications for geometric modeling and animation, and demonstrate its potential through the magic photo cube, a representative example of an edge-connected articulated solid.

\subsection{Motivation}

The magic photo cube provides a compelling example of how a seemingly simple geometric object can exhibit surprisingly complex dynamic behavior. It consists of eight rigid cubical blocks connected along selected edges by piano hinges, initially forming a compact \(2\times2\times2\) assembly (see Figure\ref{fig:vg_cube}). Through a sequence of rotations, the assembly can unfold and refold into different configurations, revealing previously hidden surfaces while maintaining its connectivity. Although the individual components are geometrically simple, understanding the motion of the complete assembly is challenging. Each folding operation changes the positions and orientations of multiple blocks, as well as the spatial locations of the hinge axes governing subsequent motions.

The visual transformations of the magic photo cube are easy to observe, but tracking the positions and orientations of its individual components is considerably more difficult. This becomes particularly important when different folding sequences produce compact assemblies at different positions and orientations in space. Such differences do not necessarily imply distinct internal configurations, since two assemblies may be equivalent under a single rigid transformation. Understanding these relationships requires a representation that preserves the identities of individual components, their connectivity, and their transformations throughout the folding process.

Beyond its recreational and artistic appeal, the magic photo cube is particularly interesting from a geometric modeling perspective. Its components are connected along edges rather than faces, making it a natural representative of edge-connected non-manifold cell complexes. This structure suggests an alternative to conventional articulated modeling, in which hinges are represented as separate geometric or mechanical elements. Instead, the connectivity of the solid cells can encode their hinge attachments, while geometric constraints determine their allowable relative motions. The magic photo cube therefore provides an accessible yet sufficiently complex example for investigating how non-manifold cell complexes can support the representation, animation, and analysis of articulated solids.

An important characteristic of the magic photo cube is the cyclic structure of its hinge connectivity. If each rigid cubical cell is represented as a graph vertex and each piano-hinge connection as a graph edge, the resulting connectivity graph forms a closed cycle. Unlike an open chain of articulated components, this closed-loop structure imposes additional geometric constraints: the relative rotations of neighboring cells must remain compatible throughout the assembly. Consequently, individual hinge motions cannot generally be prescribed independently, and folding requires coordinated transformations of multiple components. This relationship between the connectivity of the cells and the constraints governing their motion motivates our investigation of edge-connected non-manifold cell complexes as a unified framework for representing and animating articulated solids.

\subsection{Modeling using non-manifold cell complexes}

A central motivation for this work is our investigation of articulated structures formed by connecting rigid solids along selected vertices, edges, or faces. Such structures can be represented as non-manifold cell complexes, providing a unified description of their constituent cells and the incidence relationships between them. In particular, piano-hinged solids form a class of edge-connected structures in which rigid components share selected edges that serve as axes of relative rotation. This representation offers a natural alternative to conventional approaches that introduce hinges as separate geometric elements.

The magic photo cube provides a representative example of this class. Its eight rigid cubical cells are connected through prescribed edge attachments, forming an articulated structure whose connectivity remains unchanged throughout its folding motions. By representing these attachments as incidence relationships within a non-manifold cell complex, we can preserve the identities of individual cells and describe their relative transformations without explicitly modeling the physical hinges. This approach provides a foundation for analyzing the geometry and motion of edge-connected articulated solids.

\begin{figure}[hbtp]
\centering
\includegraphics[width=0.99\textwidth]{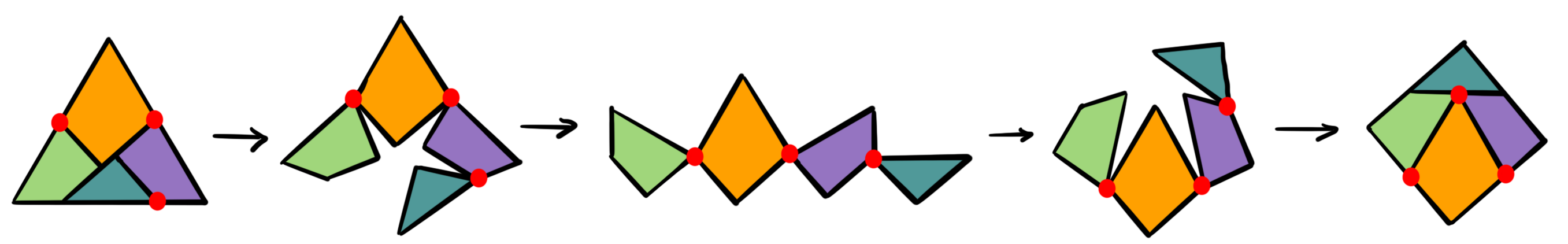}
    \caption{An example of a vertex-hinged (swing-hinged) dissection \citep{frederickson2002}. Four polygonal pieces remain connected at selected vertices, indicated by red points, while undergoing a sequence of planar rotations. The initial triangular arrangement is progressively unfolded and rearranged into a different compact configuration. Throughout the transformation, the identities of the pieces and their vertex connections remain unchanged, even though their relative positions and orientations change.}
\label{fig_swinghinged_0}
\end{figure}

\subsection{Animation using non-manifold cell complexes}

Representing articulated solids as non-manifold cell complexes provides a natural foundation for their animation. Each rigid cell maintains its own geometric identity and local coordinate system, while the incidence relationships between cells remain unchanged throughout the motion. Animation can therefore be described through rigid transformations of individual cells, subject to geometric constraints imposed by their shared edges. Rather than treating hinges as independent geometric objects, the transformations are determined by the attachment relationships encoded in the cell complex.

An important advantage of this approach is its ability to describe coordinated motions in closed-loop articulated structures. In the magic photo cube, the eight cubical cells form a cyclic connectivity graph, requiring the transformations of neighboring cells to remain compatible. Rotating one group of cells changes the spatial positions and orientations of other cells and their associated hinge axes. Consequently, subsequent rotations must be evaluated using the updated geometric configuration while preserving the original connectivity. This separation between fixed topological relationships and changing geometric transformations provides a consistent framework for constructing and analyzing folding sequences.

The resulting animation is more than a visual simulation. By tracking the transformations of labeled cells, we can compare different folding sequences, identify admissible configurations, and determine whether apparently different final assemblies are equivalent under a global rigid transformation. The magic photo cube demonstrates these capabilities through two distinct folding routes that lead to rigidly equivalent terminal configurations.

\subsection{Contributions}

The primary contribution of this paper is to demonstrate how
piano-hinged solids can be represented and animated as a
particular class of edge-connected non-manifold cell complexes.
Using the magic photo cube as a representative example,
we investigate how the incidence relationships between rigid
cells can provide a unified foundation for geometric modeling,
articulated motion, and configuration analysis.

More specifically, our contributions are as follows:

\begin{enumerate}

\item \textbf{Non-manifold representation of piano-hinged solids.}
We describe a representation in which rigid polyhedral cells
are connected through identified edges, allowing hinge
attachments to be encoded directly through cell incidence
relationships without explicitly modeling the physical hinges.
The representation preserves the identities of individual
cells and their connectivity throughout articulated motion.

\item \textbf{Animation of edge-connected articulated structures.}
We demonstrate how rigid transformations can be applied to
individual cells and groups of cells while maintaining their
prescribed hinge attachments. The separation of fixed
topological connectivity from changing geometric configurations
provides a consistent framework for describing folding
operations and tracking moving hinge axes.

\item \textbf{Analysis of cyclic connectivity and coordinated motion.}
We interpret the magic photo cube as a closed chain of eight
rigid cells connected by eight edge hinges. Its cyclic
connectivity imposes geometric compatibility constraints
that couple the motions of its constituent cells.
We demonstrate how these constraints are reflected in the
coordinated folding operations of the cube.

\item \textbf{Reconstruction and comparison of folding sequences.}
We reconstruct two distinct three-stage folding sequences
of the magic photo cube and track the positions and
orientations of its eight labeled cells. We show that,
despite producing terminal assemblies at different positions
and orientations in space, the resulting configurations
are rigidly equivalent under a single proper rigid
transformation.

\item \textbf{Interactive geometric visualization.}
We provide an interactive implementation that supports
the visualization of folding sequences, individual cell
transformations, hinge connectivity, and surface exposure.
The implementation also supports assigning images to the
twelve available picture panels, illustrating the relationship
between the underlying geometric structure and the cube's
artistic applications. The interactive geometric and textured demonstrations are publicly available through the accompanying project website~\cite{Akleman2026MagicCubeWebsite}.

\end{enumerate}

\section{Previous Work}\label{sec:previous-work}

The study of magic cubes and hinged structures spans several distinct areas of mathematics, computational science, and engineering. Although the term \emph{magic cube} is used in different contexts, it can refer to substantially different mathematical and physical objects. We first distinguish these interpretations from the articulated magic photo cube investigated in this paper and introduce existing examples of this particular mechanism. We then review the geometric foundations of hinged dissections, particularly Frederickson's work on vertex-hinged and piano-hinged structures, followed by computational research on hinged transformations. Finally, we identify the geometric and topological characteristics that distinguish the magic photo cube from conventional hinged dissections and motivate its representation as an edge-connected non-manifold cell complex.

\begin{figure}[hbtp]
\centering
\includegraphics[width=0.99\textwidth]{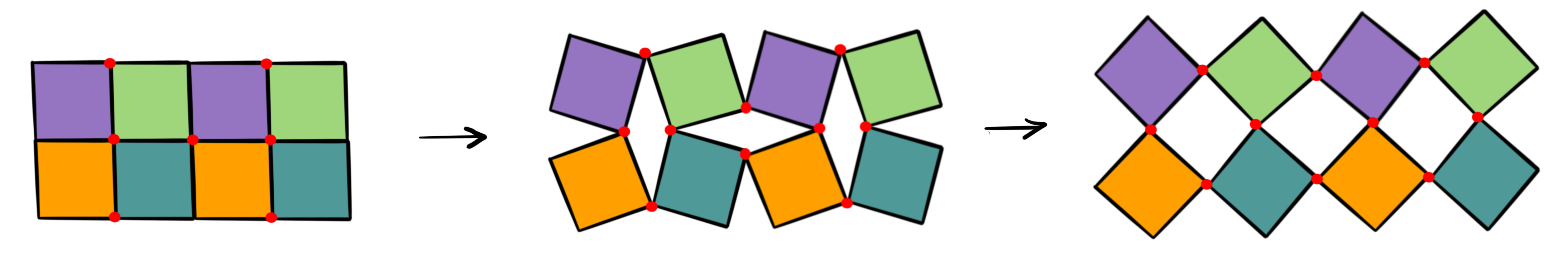}
    \caption{An example of a vertex-hinged auxetic mechanism consisting of rigid square elements connected at selected vertices \citep{GrimaEvans2000}. Coordinated rotations of the squares transform the assembly from a compact configuration into a more open arrangement, producing expansion in multiple directions. Such rotating-unit mechanisms are closely related to auxetic structures, which can exhibit a negative Poisson's ratio under suitable deformation conditions \citep{schneider2024concurrent, ebert2024abc, schneider2025topology}. The red points indicate the persistent hinge connections between adjacent elements. Throughout the transformation, the individual pieces remain rigid, while their relative orientations and the overall dimensions of the assembly change.}
\label{fig_swinghinged_1}
\end{figure}

\subsection{Different Interpretations of Magic Cubes}

The term \emph{magic cube} has been used in several distinct mathematical, computational, and engineering contexts. In classical recreational mathematics, magic cubes are three-dimensional arrangements of integers satisfying prescribed summation conditions along rows, columns, and diagonals. \citep{Trenkler2000MagicCubes}, for example, investigated mathematical constructions of such numerical arrangements.

The term has also appeared in computational applications involving information representation and transformation. \citep{Rani2022FusedMagicCube} introduced fused magic cubes for image encryption, using their numerical structure to support data permutation and encryption operations. In a different application, \citep{Chu2025MagicCubeMetamaterials} investigated mechanically reconfigurable magic-cube metamaterials for controlling electromagnetic responses. Although these studies demonstrate the broad use of the term, their objectives and underlying mathematical structures differ from those investigated here.

\begin{figure}[hbtp]
\centering
\includegraphics[width=0.99\textwidth]{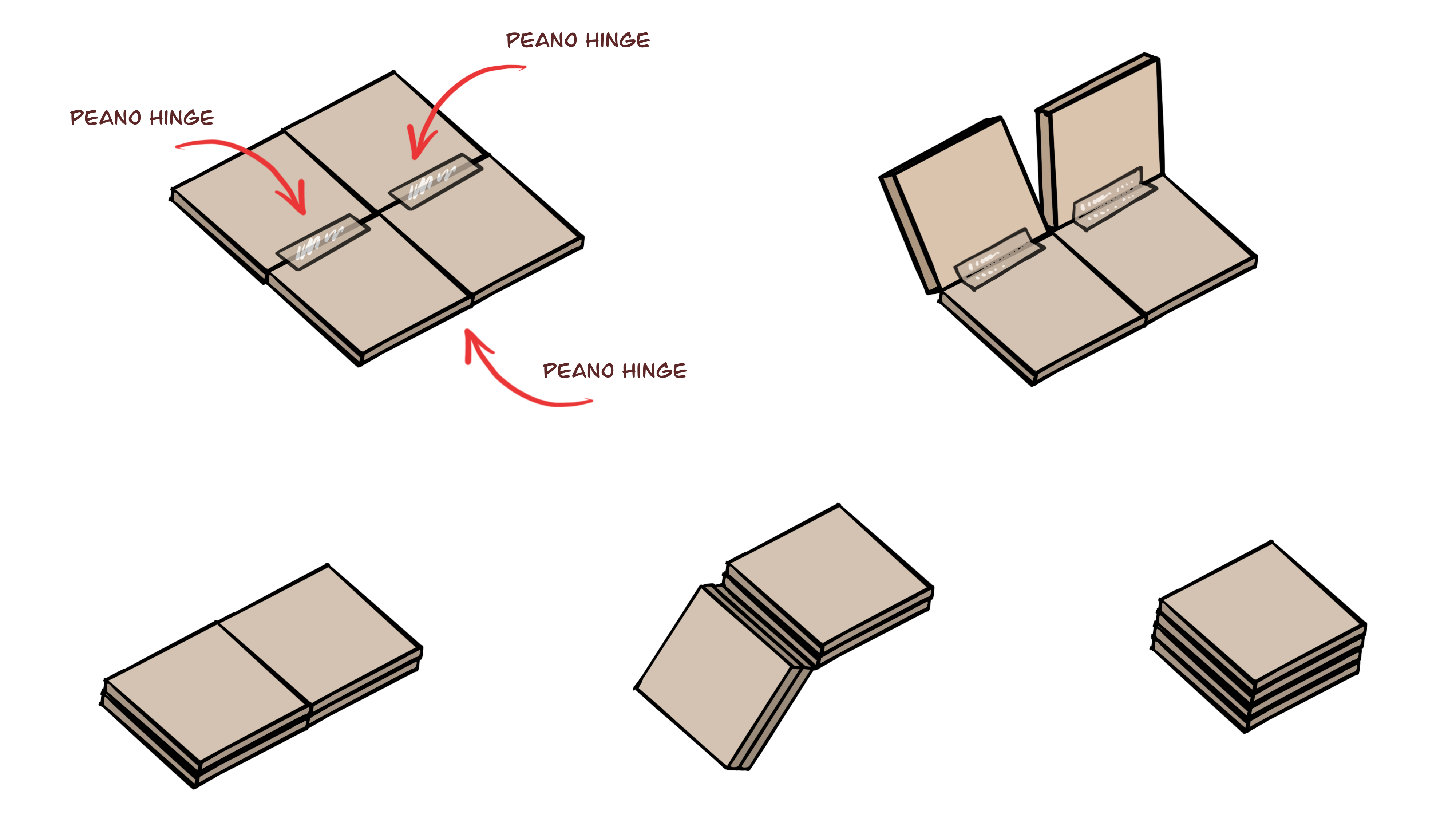}
    \caption{An example of a piano-hinged dissection formed
from thin rectangular solid panels connected along selected
edges. The sequence illustrates how the panels rotate about
their hinge axes and fold into a compact stack. Unlike
vertex-hinged planar dissections, piano-hinged constructions
permit out-of-plane rotations and stacking of the constituent
solids. The connections remain attached to the same material
edges throughout the transformation, while the spatial
positions and orientations of the panels change.}
\label{fig_images_Peano_hinged_0}
\end{figure}

Our interest is specifically in the \emph{magic photo cube}, an articulated assembly of eight rigid cubical cells connected by piano hinges along selected edges. Commercial examples, including a Van Gogh-themed magic photo cube~\citep{vangogh_cube}, demonstrate how such assemblies can be folded to reveal different sets of images. The Van Gogh Museum's historical documentation also refers to a Magic Cube among its museum-shop products~\citep{vgm_2003}. Practical construction demonstrations by~\citep{chica_video,tanu_video} illustrate the folding process and the importance of maintaining consistent image orientations across the connected components.

Unlike numerical magic cubes, the essential properties of the magic photo cube arise from the connectivity and relative motions of its constituent solids. Unlike applications primarily concerned with material responses or information encoding, our objective is to investigate the geometric and topological representation of the articulated structure itself. Existing commercial examples and construction demonstrations emphasize the visual transformations and image-display capabilities of the cube, whereas our investigation focuses on the identities, connectivity, and spatial transformations of its individual rigid cells.

An important characteristic of the magic photo cube is that its hinge-connectivity graph forms a closed cycle. This cyclic structure introduces geometric compatibility constraints that couple the motions of its constituent cells. Consequently, the magic photo cube provides a representative example of an edge-connected articulated structure whose geometric behavior can be investigated through a non-manifold cell-complex representation.

\subsection{Geometric Dissections and Hinged Structures}

Geometric dissection studies the decomposition of a figure into finitely many rigid pieces that can be rearranged into another figure. In the plane, the Wallace--Bolyai--Gerwien theorem provides the classical equidecomposability result for equal-area polygons. A \emph{hinged dissection} imposes a stronger requirement: the pieces remain connected during reconfiguration. Thus, the problem involves not only geometric decomposition but also the kinematics and possible interference of connected rigid pieces.

\begin{figure}[hbtp]
\centering
\includegraphics[width=0.99\textwidth]{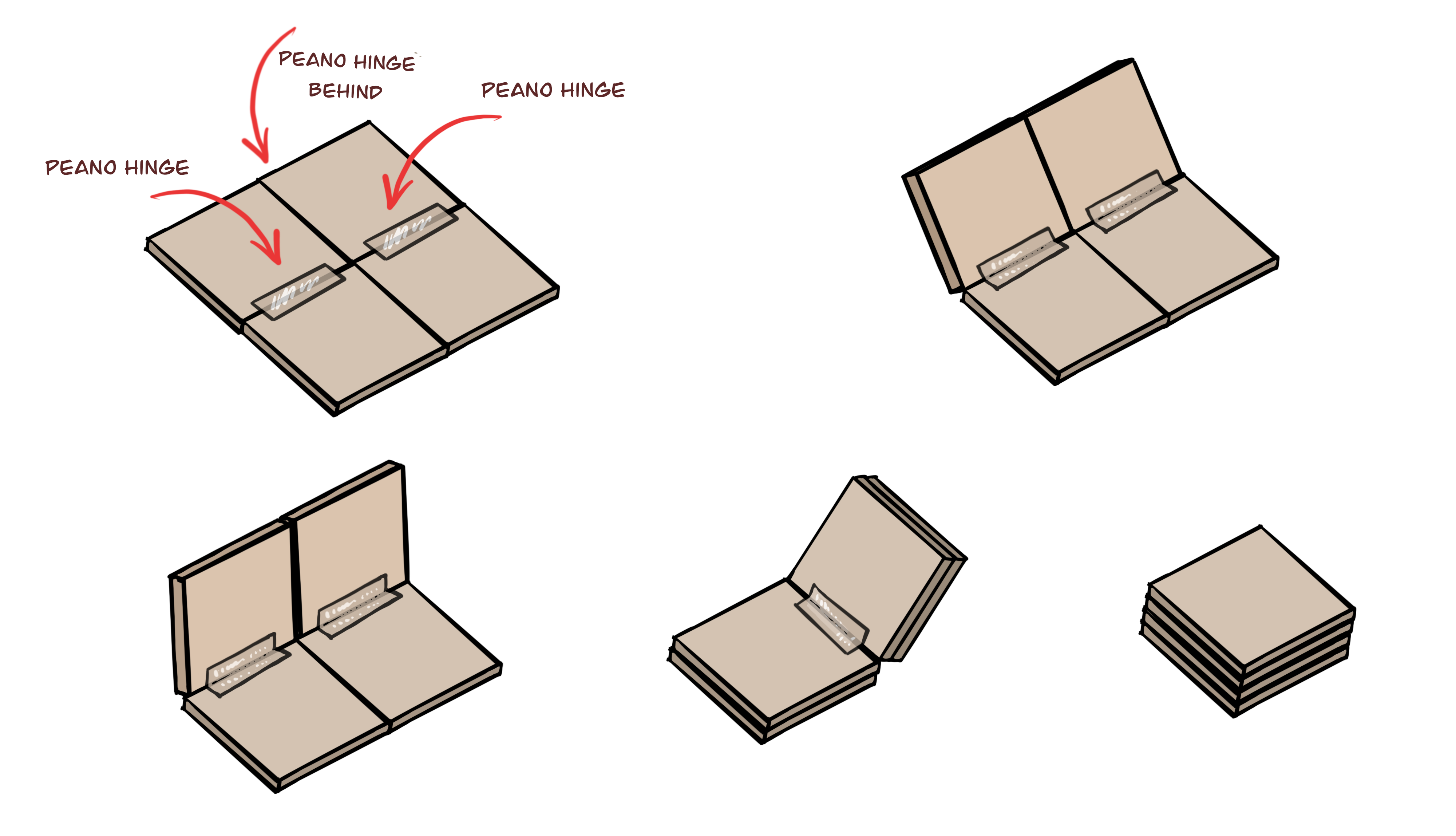}
    \caption{Five stages in the folding of a piano-hinged assembly
of thin rectangular panels. The initial configuration shows
the arrangement of the panels and their edge hinges, including
connections located on opposite sides of the assembly.
Successive rotations bring the panels into partially folded
configurations, followed by the formation of a compact stack.
The final configuration consists of the original panels
superimposed in a layered arrangement. This example
demonstrates the importance of specifying the exact edges
participating in each hinge connection: the geometric
behavior of the assembly depends on these edge
identifications, not merely on which components are
connected.}
\label{fig_images_Peano_hinged_1}
\end{figure}

Frederickson developed a systematic collection of constructions and examples in \emph{Dissections: Plane and Fancy}, \emph{Hinged Dissections: Swinging and Twisting}, and \emph{Piano-Hinged Dissections: Time to Fold!}~\citep{frederickson1997,frederickson2002,frederickson2006}. These works provide an important geometric foundation for understanding how rigid pieces can remain connected while being reconfigured into different shapes.

Two types of hinge connections are particularly relevant to these constructions. In vertex-hinged (or swing-hinged) dissections, rigid pieces are connected at selected vertices, allowing relative rotations that transform one planar figure into another (see Figures~\ref{fig_swinghinged_0} and~\ref{fig_swinghinged_1}). In piano-hinged dissections, connections extend along edges, permitting relative rotations about hinge axes (see Figures~\ref{fig_images_Peano_hinged_0} and~\ref{fig_images_Peano_hinged_1}). The latter constructions often involve folding or stacking thin solid pieces over one another rather than rearranging them entirely within a common plane.

A distinguishing characteristic of many classical hinged-dissection constructions is their predominantly planar geometry. Although physical realizations may have a small but nonzero thickness, their geometric organization is largely determined by two-dimensional figures. Such structures may be viewed as essentially \emph{2.5-dimensional} articulated solids. Piano-hinged constructions introduce out-of-plane rotations and stacking, but the underlying pieces are still typically derived from planar dissections.

Hinged dissections have also been studied algorithmically. Demaine and collaborators examined hinged constructions for polyominoes and related polyforms~\citep{demaine2005polyominoes}. Abbott and collaborators proved a general existence theorem for common hinged dissections of finite collections of equal-area polygons~\citep{abbott2012hinged}. These results demonstrate that hinged dissections constitute a substantial mathematical problem extending beyond individual recreational constructions. They also motivate broader questions concerning the connectivity, allowable transformations, and possible interference of articulated rigid pieces. The present work does not seek a comparable general existence theorem; instead, it investigates the representation and animation of a particular class of edge-connected articulated solids.

An important distinction between many classical hinged dissections and the magic photo cube concerns the topology of their connectivity graphs. Hinged-dissection constructions frequently employ open chains of connected pieces, in which relative rotations can be considered without imposing a global loop-closure condition. In contrast, the eight cubical cells of the magic photo cube form a closed cycle through their hinge connections. As a result, the relative transformations of neighboring cells must satisfy geometric compatibility constraints around the entire cycle. Individual hinge rotations cannot generally be prescribed independently, and the folding process requires coordinated motions of multiple components.

The magic photo cube also differs from predominantly planar hinged dissections in its genuinely three-dimensional organization. Its components are volumetric cubes rather than thin extrusions of planar figures, and their motions involve changing spatial positions, orientations, and hinge axes. Nevertheless, Frederickson's constructions and the subsequent computational literature provide an important conceptual foundation for our investigation. Both classical hinged dissections and the magic photo cube preserve the connectivity of rigid components during reconfiguration. The distinguishing objective of the present work is to represent these connections directly through the incidence relationships of a non-manifold cell complex and to investigate how this representation supports the animation and configuration analysis of closed-loop, edge-connected articulated solids.

\section{Theoretical Framework}\label{sec:theoretical-framework}

We investigate a class of articulated structures constructed by connecting otherwise manifold surface meshes along selected edges. Each component is a rigid solid bounded by a closed two-manifold surface, while the complete assembly is represented by a non-manifold cell complex obtained through prescribed edge identifications. The resulting structure separates the topology of the individual components from the topology of their connections. This distinction provides a natural foundation for representing piano-hinged solids and analyzing their allowable motions.

\subsection{Edge-Connected Non-Manifold Surface Complexes}
\label{sec:edge-connected-complexes}

We consider a particular class of non-manifold complexes constructed by connecting otherwise manifold surface meshes along selected edges. Each component represents the boundary of a rigid solid, and each connection is defined by identifying two specific edges belonging to two distinct component meshes. Unlike representations that introduce hinges as separate geometric objects, the proposed construction incorporates the hinge connections directly into the incidence relationships of the resulting non-manifold complex.

Consider a finite collection of rigid solids
\begin{equation}
\mathcal{S}=\{S_1,S_2,\ldots,S_n\},
\end{equation}
where each solid $S_i$ is bounded by a closed, orientable, polygonal two-manifold mesh $M_i=\partial S_i$. Each mesh consists of vertices, edges, and polygonal faces satisfying the usual manifold incidence relationships. In particular, every edge of a closed two-manifold mesh is incident to exactly two faces.

We construct an articulated assembly by identifying selected edges belonging to different component meshes. Let
\begin{equation}
\mathcal{H}=\{h_1,h_2,\ldots,h_m\}
\end{equation}
denote the collection of prescribed edge identifications. Each connection is represented by
\begin{equation}
h_k=(e_i,e_j,\phi_k),
\end{equation}
where $e_i\subset M_i$ and $e_j\subset M_j$ are specific edges of two distinct manifold meshes, and $\phi_k:e_i\rightarrow e_j$ is a homeomorphism defining their correspondence. For polygonal edges, this correspondence is determined by matching their endpoints, with either orientation permitted. For a realizable rigid piano hinge, the corresponding edges must also have equal lengths.

The resulting complex is defined as the quotient
\begin{equation}
K=\left(\bigsqcup_{i=1}^{n}M_i\right)\Big/\sim_{\mathcal{H}},
\end{equation}
where $\bigsqcup$ denotes the disjoint union of the component meshes and $\sim_{\mathcal{H}}$ is the equivalence relation generated by the prescribed edge identifications. More precisely, for every $h_k=(e_i,e_j,\phi_k)$,
\begin{equation}
x\sim_{\mathcal{H}}\phi_k(x),
\qquad x\in e_i.
\end{equation}

This construction preserves the identities of the original manifold components while establishing their shared edge incidences. The complete representation therefore specifies not only which components are connected, but also precisely which edges are identified and how their endpoints correspond. These identifications are intrinsic to the non-manifold complex and do not require an additional description of hinge placement.

Although every component $M_i$ is individually a two-manifold, the resulting complex $K$ is generally non-manifold. Consider a connection between two closed manifold meshes along a single edge. Before identification, each edge is incident to two faces of its own manifold. After identification, the shared edge is incident to four faces, two from each component. Consequently, a neighborhood of an interior point of the identified edge is not homeomorphic to an open disk, as required for a two-manifold.

The non-manifold structure is therefore localized at the identified edges and their incident vertices, while the remaining portions of the component surfaces retain their original manifold neighborhoods. This gives a restricted and geometrically meaningful class of non-manifold complexes constructed from manifold components through explicit edge identifications.

For articulated solid modeling, we retain the corresponding three-dimensional solids $S_i$ together with their boundary meshes. The complex $K$ describes the topological incidence relationships between their boundaries, while the solids provide the rigid volumetric components of the assembly. The identification of two edges specifies their persistent connection, and the corresponding geometric realization constrains those edges to occupy the same spatial segment throughout articulation.

An important consequence is that the non-manifold complex contains more information than an ordinary connectivity graph. A graph can indicate that two rigid components are connected, but it does not specify which edges of their boundary meshes are identified. For example, connecting a sequence of rectangular panels along consistently positioned hinge edges produces a different edge-incidence structure from connecting the same panels with alternating hinge placements. Although both assemblies may have the same abstract connectivity graph, their complete non-manifold representations distinguish the two constructions.

Thus, the proposed representation separates three related concepts: the manifold geometry of the individual rigid components, the non-manifold incidence structure defined by their edge identifications, and the geometric realization of the articulated assembly. The incidence structure remains fixed during motion, while the positions and orientations of the constituent solids change subject to the constraints imposed by their identified edges. In the following subsection, we classify these structures according to the connectivity graphs induced by their edge identifications.

\subsection{Connectivity Graphs and Their Classification}\label{sec:connectivity-graphs}

The edge-identification structure can be represented by a graph
\begin{equation}
G=(V,E),
\end{equation}
where each graph vertex $v_i\in V$ corresponds to a rigid component $S_i$, and each graph edge $(v_i,v_j)\in E$ represents a prescribed piano-hinge connection between components $S_i$ and $S_j$. When two components share more than one independently specified hinge connection, the representation may be treated as a multigraph.

The connectivity graph is a simplified representation derived from the non-manifold cell complex. It records which components are connected but does not retain the specific edges involved in each connection. In the complete non-manifold representation, every hinge is defined by the identification of two particular edges belonging to distinct manifold meshes, together with their endpoint correspondence. Thus, the precise hinge locations are intrinsic to the non-manifold cell complex, rather than additional information associated with the graph.

The graph $G$ captures the connectivity of the articulated assembly independently of the geometric shapes of its components. In particular, the same connectivity graph can describe assemblies constructed from cubes, prisms, or more general polyhedral solids, provided that compatible hinge edges can be selected.

Several classes of connectivity graphs are especially relevant.

\paragraph{Open chains.}
An open chain is represented by a path graph $P_n$, consisting of $n$ vertices connected by $n-1$ edges. Its two terminal vertices have degree one, while all intermediate vertices have degree two. Such structures are common in articulated mechanisms and hinged dissections. Their connectivity contains no closed loops, so successive relative hinge transformations can be propagated from one component to the next without imposing a global loop-closure condition.

\paragraph{Trees.}
A tree is a connected graph containing no cycles. It generalizes an open chain by allowing branching structures in which a component may connect to three or more neighboring components. For a tree with $n$ vertices, the number of connections is $n-1$. Selecting one component as a root establishes a hierarchy through which the transformations of the remaining components can be determined from their relative hinge transformations. Although branching introduces additional geometric interactions and possible collisions, the connectivity itself does not impose loop-closure constraints.

\paragraph{Closed cycles.}
A closed chain is represented by a cycle graph $C_n$, in which every vertex has degree two and the number of connections equals the number of components. Unlike an open chain, a closed cycle contains a global compatibility condition: the relative transformations accumulated around the cycle must return consistently to the initial component. Consequently, hinge rotations cannot generally be assigned independently.

\paragraph{General graphs.}
More general articulated assemblies may have connectivity graphs containing branches, multiple cycles, or combinations of tree-like and cyclic substructures. The presence of multiple independent cycles introduces additional loop-closure conditions. Such graphs provide a general description of edge-connected articulated solids without restricting their organization to serial chains or a single closed loop.

For a connected graph, the number of independent cycles is given by its cycle rank,
\begin{equation}
\beta_1(G)=|E|-|V|+1.
\end{equation}
Thus, an open chain or tree has $\beta_1(G)=0$, a simple closed cycle has $\beta_1(G)=1$, and a connected graph with multiple independent cycles has $\beta_1(G)>1$. This quantity characterizes the topology of the connectivity graph, although it does not by itself determine the mobility of the corresponding geometric mechanism.

The distinction between graph connectivity and edge incidence is important for geometric modeling. Two assemblies may have isomorphic connectivity graphs while possessing different
non-manifold cell complexes because their edge identifications differ. For example, six rectangular panels connected in an open chain can form an open hexagonal prism when their hinge edges are consistently arranged, whereas alternating hinge placements can permit the panels to fold into a compact stack. Both assemblies have connectivity graph $P_6$, but their complete edge-incidence structures distinguish the two mechanisms. Consequently, the connectivity graph classifies the organization of the assembly, while the non-manifold cell
complex specifies its actual connections.

\subsection{Geometric Realization and Articulation}\label{sec:geometric-realization}

The abstract complex $K$ specifies which components are connected, but does not determine their positions or orientations in three-dimensional space. To describe an articulated configuration, we assign a rigid transformation to each component.

Let
\begin{equation}
T_i=(R_i,\mathbf{t}_i)\in SE(3),
\end{equation}
where $R_i\in SO(3)$ is the orientation of component $S_i$ and $\mathbf{t}_i\in\mathbb{R}^3$ is its translation. A point $\mathbf{x}$ expressed in the local coordinates of component $i$ is mapped into world coordinates by
\begin{equation}
T_i(\mathbf{x})=R_i\mathbf{x}+\mathbf{t}_i.
\end{equation}

A configuration of the articulated assembly is therefore represented by
\begin{equation}
\mathcal{C}=(T_1,T_2,\ldots,T_n)\in SE(3)^n.
\end{equation}

The topology of the complex remains fixed during articulation. Only the transformations of its constituent components change. In particular, the vertex, edge, and face identities of each component remain attached to that component throughout the motion.

Consider a hinge connection between components $S_i$ and $S_j$. Let its corresponding local edge endpoints be $\mathbf{a}_i,\mathbf{b}_i$ on component $i$ and $\mathbf{a}_j,\mathbf{b}_j$ on component $j$. The geometric realization must satisfy
\begin{equation}
T_i(\mathbf{a}_i)=T_j(\mathbf{a}_j),\qquad
T_i(\mathbf{b}_i)=T_j(\mathbf{b}_j).
\end{equation}

These conditions ensure that the identified edges occupy the same geometric segment in world coordinates. They allow the two components to rotate relative to one another about that segment while preventing separation of the hinge endpoints.

The world-space hinge axis is therefore not fixed independently of the components. It is transported by their rigid transformations. If a hinge has local endpoints $\mathbf{a}_i$ and $\mathbf{b}_i$, its current axis is the line through $T_i(\mathbf{a}_i)$ and $T_i(\mathbf{b}_i)$. This provides a direct geometric interpretation of piano-hinge motion without introducing a separate solid model of the hinge itself.

\subsection{Hinge Constraints and Degrees of Freedom}\label{sec:hinge-constraints}

The connectivity graph describes the combinatorial organization of the assembly, while geometric constraints determine its admissible configurations. For every hinge connection, the corresponding edge endpoints must coincide after applying the transformations of the incident components.

Let $\mathcal{Q}$ denote the set of configurations satisfying all hinge constraints:
\begin{equation}
\mathcal{Q}=\left\{\mathcal{C}\in SE(3)^n:
T_i(\mathbf{a}_i)=T_j(\mathbf{a}_j),\
T_i(\mathbf{b}_i)=T_j(\mathbf{b}_j)
\text{ for every }(i,j)\in E
\right\}.
\end{equation}

The space $\mathcal{Q}$ describes kinematic compatibility, but does not necessarily exclude collisions between nonadjacent components. Additional nonpenetration conditions are required when physical realizability is considered.

For a single piano hinge connecting two rigid solids, the relative motion is a rotation about the common hinge axis, giving one rotational degree of freedom. In an open chain containing $n$ rigid components and $n-1$ independent hinges, fixing the pose of one component leaves $n-1$ independent hinge-angle parameters, before considering collision restrictions.

The same reasoning extends to tree-connected assemblies. Because a tree contains no cycles, each hinge introduces one relative rotational parameter, and the transformations of all components can be propagated from a selected root. Consequently, an ideal tree-connected assembly with $n$ components has $n-1$ internal rotational degrees of freedom, provided that the hinge constraints are geometrically consistent and no additional constraints are imposed.

Closed cycles behave differently. Consider a cycle containing $n$ rigid components and $n$ hinges. Let $H_k(\theta_k)\in SE(3)$ denote the relative rigid transformation associated with hinge $k$, including its fixed geometric attachment frames and its rotational parameter $\theta_k$. Consistency around the cycle requires
\begin{equation}
H_1(\theta_1)H_2(\theta_2)\cdots H_n(\theta_n)=I,
\end{equation}
where $I$ is the identity rigid transformation.

This equation expresses the loop-closure condition. It imposes coupled restrictions on the hinge angles, and the number of independent rotational parameters depends on the particular geometry of the mechanism. In general, the mobility of a closed-chain mechanism cannot be determined from its connectivity graph alone.

For general graphs containing multiple independent cycles, corresponding closure conditions must be satisfied for a basis of independent graph cycles. Although the number of such cycles is given by $\beta_1(G)$, the independence and rank of the resulting geometric constraints depend on the hinge locations, orientations, and configurations. Thus, graph topology determines the organization of the constraints, while geometry determines the actual mobility of the articulated structure.

\subsection{The Magic Photo Cube as a Cyclic Non-Manifold Complex}\label{sec:magic-cube-complex}

The magic photo cube provides a representative example of the preceding construction. Its eight rigid cubical components have closed two-manifold boundary meshes, and selected edges are identified to establish eight piano-hinge connections. The resulting assembly is an edge-connected non-manifold complex whose hinge-connectivity graph is the cycle
\begin{equation}
G_{\mathrm{cube}}\cong C_8.
\end{equation}

Since the graph contains eight vertices and eight edges, its cycle rank is
\begin{equation}
\beta_1(G_{\mathrm{cube}})=8-8+1=1.
\end{equation}

The assembly therefore contains one independent connectivity cycle. Its hinge rotations must satisfy a global closure condition, distinguishing its motion from that of an open chain or tree-connected structure. The existence of this cycle does not determine the number of available folding motions; these depend on the spatial arrangement of the hinge axes and the geometry of the eight cubical components.

The non-manifold representation provides two complementary descriptions of the mechanism. The abstract complex records the persistent identities of the cubical cells and their edge-incidence relationships, while the geometric configuration records their changing positions and orientations. During folding, the incidence relationships remain unchanged, even as the connected cells move into different spatial arrangements.

This separation is particularly useful for analyzing different folding sequences. Two configurations may appear different because their assemblies occupy different positions or orientations in a fixed reference frame. However, they represent the same labeled geometric arrangement if a single proper rigid transformation maps each component pose in one configuration to the corresponding pose in the other. Formally, configurations $\mathcal{C}^{A}$ and $\mathcal{C}^{B}$ are rigidly equivalent if there exists $T\in SE(3)$ such that
\begin{equation}
T_i^{B}=T\,T_i^{A}
\qquad \text{for all }i=1,\ldots,n.
\end{equation}

The following sections apply this framework to the magic photo cube's specific hinge arrangement, reconstruct its folding operations, and compare the resulting configurations. Our objective is not to provide a general solver for arbitrary non-manifold articulated complexes, but to demonstrate how their topology and geometry can be represented consistently and used for articulated modeling and animation.

\section{Mechanism and notation}
\subsection{Eight persistent labels}
Let $\mathcal I=\{0,1\}^3$. A block with identity $i=(i_x,i_y,i_z)$ is labeled by the bit string $i_xi_yi_z$. Its initial center is
\begin{equation}\label{eq:init}
\mathbf c_i^0=(i_x-\tfrac12,\ i_y-\tfrac12,\ i_z-\tfrac12).
\end{equation}
The identity is attached permanently to the material block, not to its instantaneous position. For visual separation the renderer uses a small adjustable gap $g$; for numerical comparisons reported here $g=0.03$.

A configuration is the collection of poses
\begin{equation}\label{eq:pose}
\mathcal C=\{(\mathbf c_i,Q_i):i\in\mathcal I\},\qquad Q_i\in SO(3),
\end{equation}
where $Q_i$ maps the local coordinate frame of block $i$ to world coordinates. A drawing of the block centers alone discards the $Q_i$ data and therefore cannot determine which decorated faces are visible.

\subsection{Edge hinges and rigid groups}
\begin{figure}[t]
\centering
\begin{subfigure}[t]{.485\textwidth}
    \centering    \includegraphics[width=.45\linewidth]{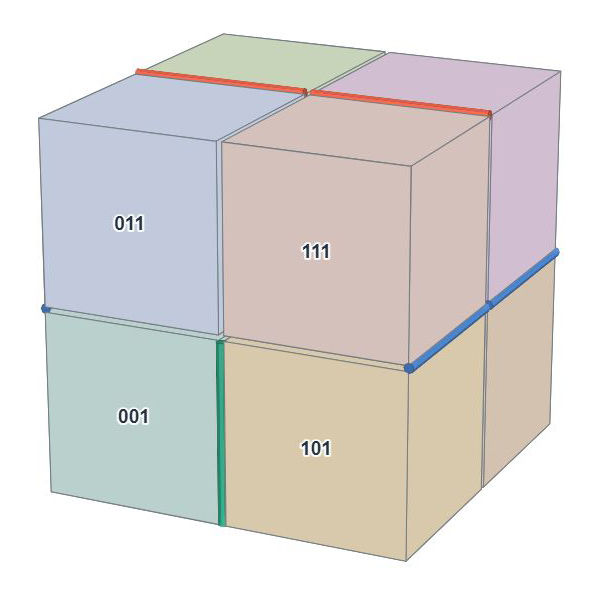}    \includegraphics[width=.45\linewidth]{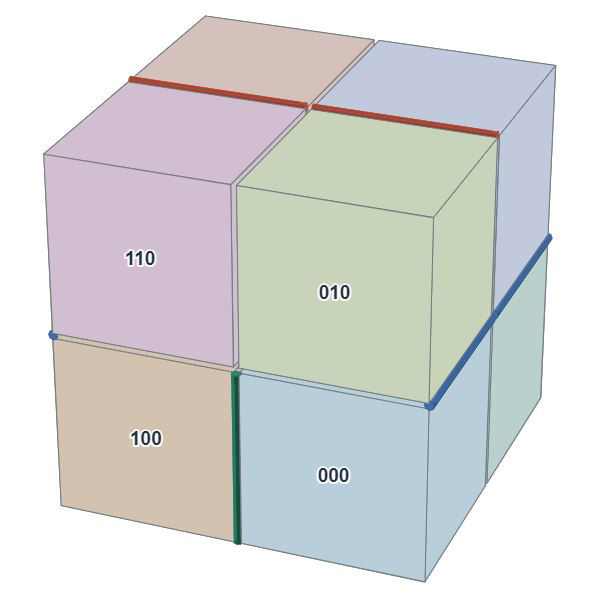}
    \caption{Two views of the initial configuration}
\end{subfigure}
\hfill
\begin{subfigure}[t]{.485\textwidth}
    \centering   \includegraphics[width=.45\linewidth]{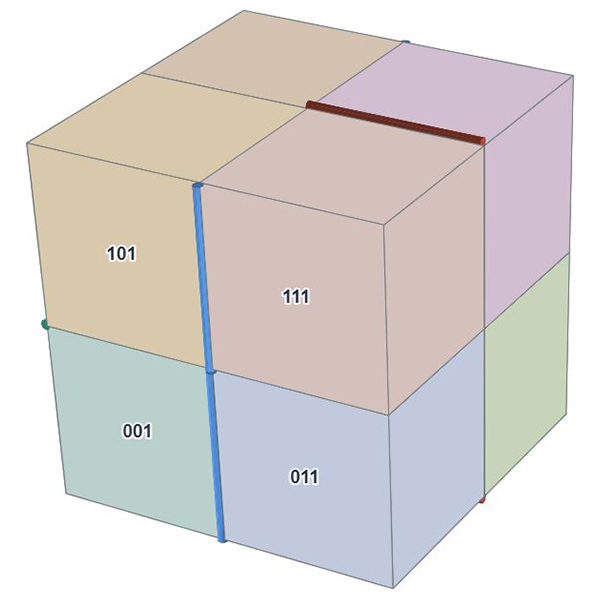}    \includegraphics[width=.45\linewidth]{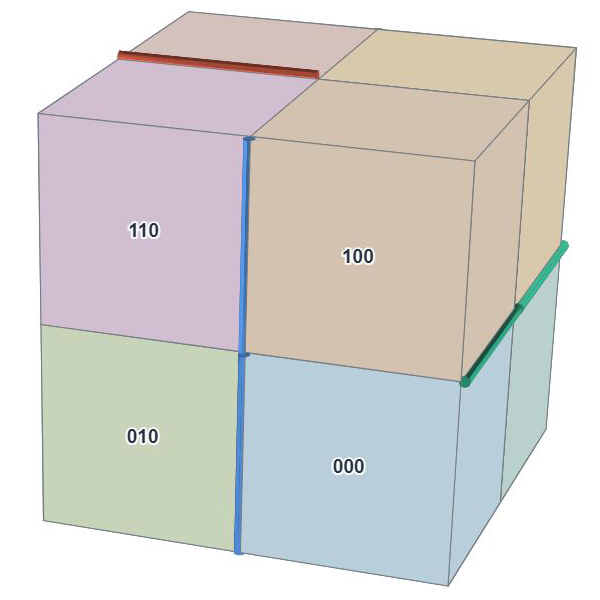}
    \caption{Two views of the final configuration}
\end{subfigure}

\caption{Initial and final eight-block magic photo cube, with persistent binary labels, colored hinge segments, and the stationary original $2\times2\times2$ reference grid. Rendered directly from the accompanying JavaScript simulation.}
\label{fig:initial}
\end{figure}

We use three hinge families, color-coded red ($R$), green ($G$), and blue ($B$). The present implementation contains two red, two green, and four blue unit-edge hinge segments. These are physical connections, not world-axis rotation commands. At each stage, a hinge line is carried into the current pose by the transformation of its attached block. Groups of blocks move rigidly about that transported line. Figure~\ref{fig:initial} shows the initial assembly and a stationary amber reference lattice.

\subsection{Rotation about a transported hinge}
For a hinge line through $\mathbf a$ in unit direction $\mathbf u$, rotation of point $\mathbf p$ through angle $\theta$ is given by Rodrigues' formula:
\begin{align}\label{eq:rodrigues}
\mathbf p'&=\mathbf a+\mathcal R(\mathbf u,\theta)(\mathbf p-\mathbf a),\\
\mathcal R(\mathbf u,\theta)\mathbf v
&=\mathbf v\cos\theta+(\mathbf u\times\mathbf v)\sin\theta
+\mathbf u(\mathbf u\cdot\mathbf v)(1-\cos\theta).
\end{align}
If stage $k$ follows stages $1,\ldots,k-1$, its axis endpoints are computed \emph{after} applying the earlier transformations. Thus, the second and third stages are not rotations about unchanging global $x,y,z$ axes. Each stage acts on complete rigid groups, and the hinge geometry must remain attached to the material blocks.

\section{Two three-stage folding routes}
We denote a compact initial state by $S_0$ and an intermediate state by its completed color sequence. A directed arrow here denotes a chosen motion segment, not an assertion that the reverse motion is forbidden.

\subsection{Route A: red--green--blue}
\begin{figure*}[t]
\centering
\begin{subfigure}{.24\textwidth}\includegraphics[width=\linewidth]{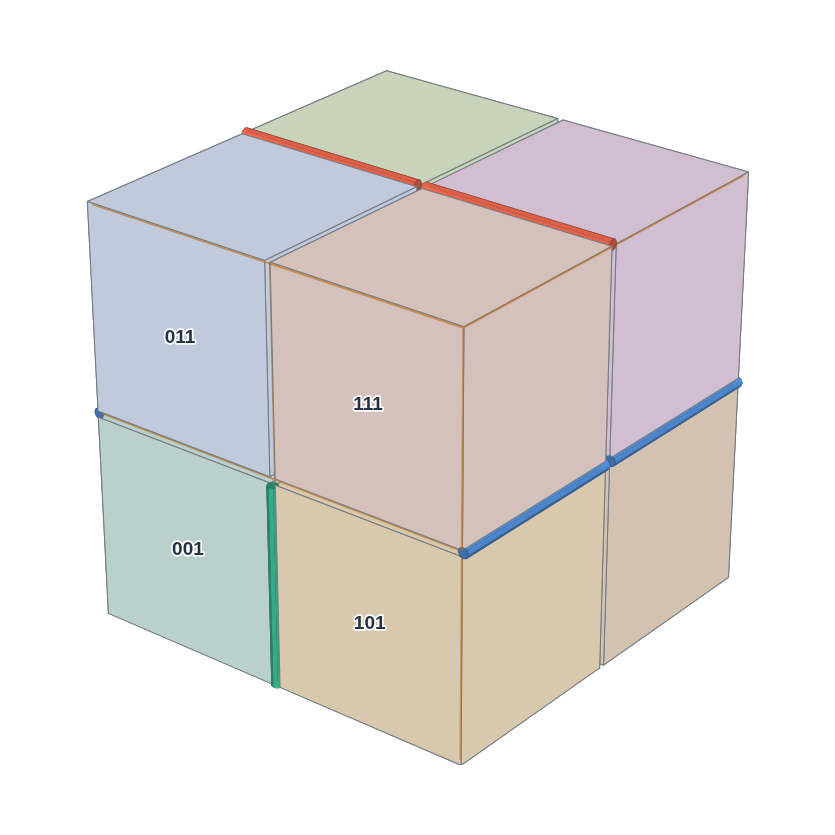}\caption{$S_0$}\end{subfigure}
\begin{subfigure}{.24\textwidth}\includegraphics[width=\linewidth]{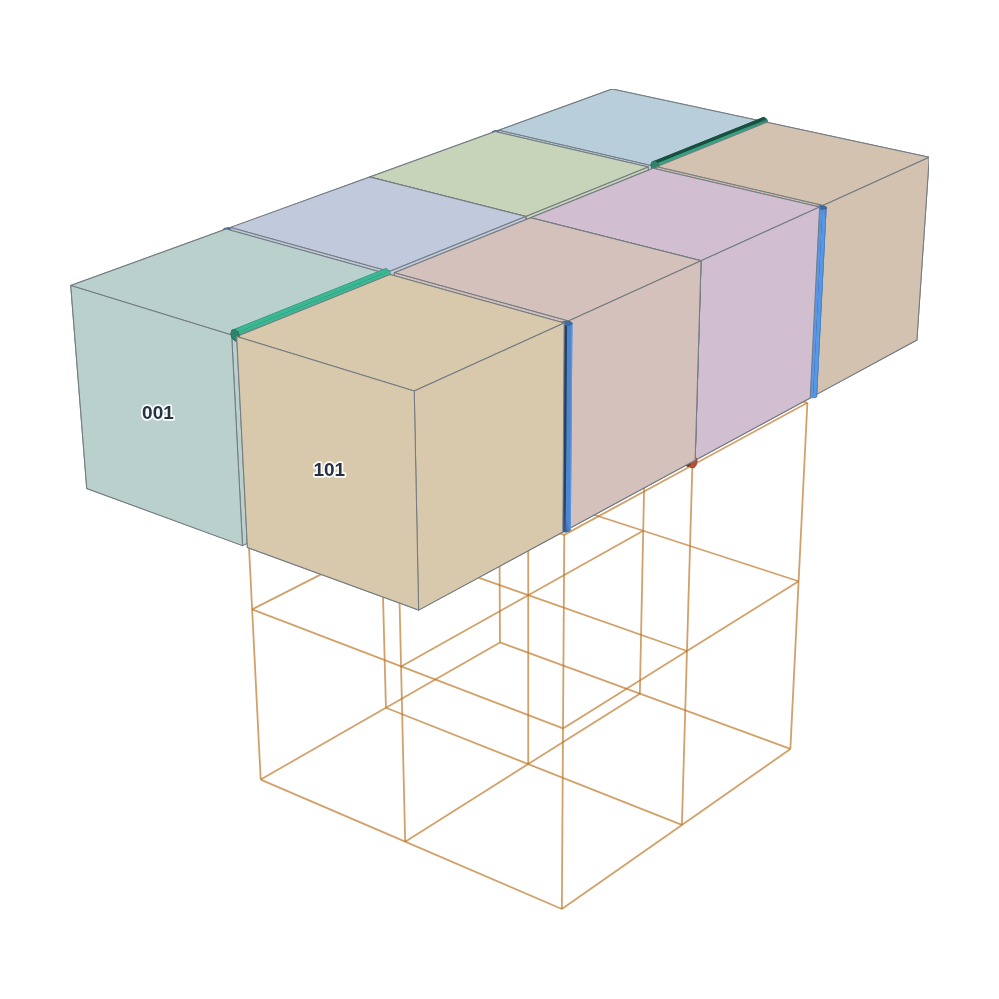}\caption{$S_R$}\end{subfigure}
\begin{subfigure}{.24\textwidth}\includegraphics[width=\linewidth]{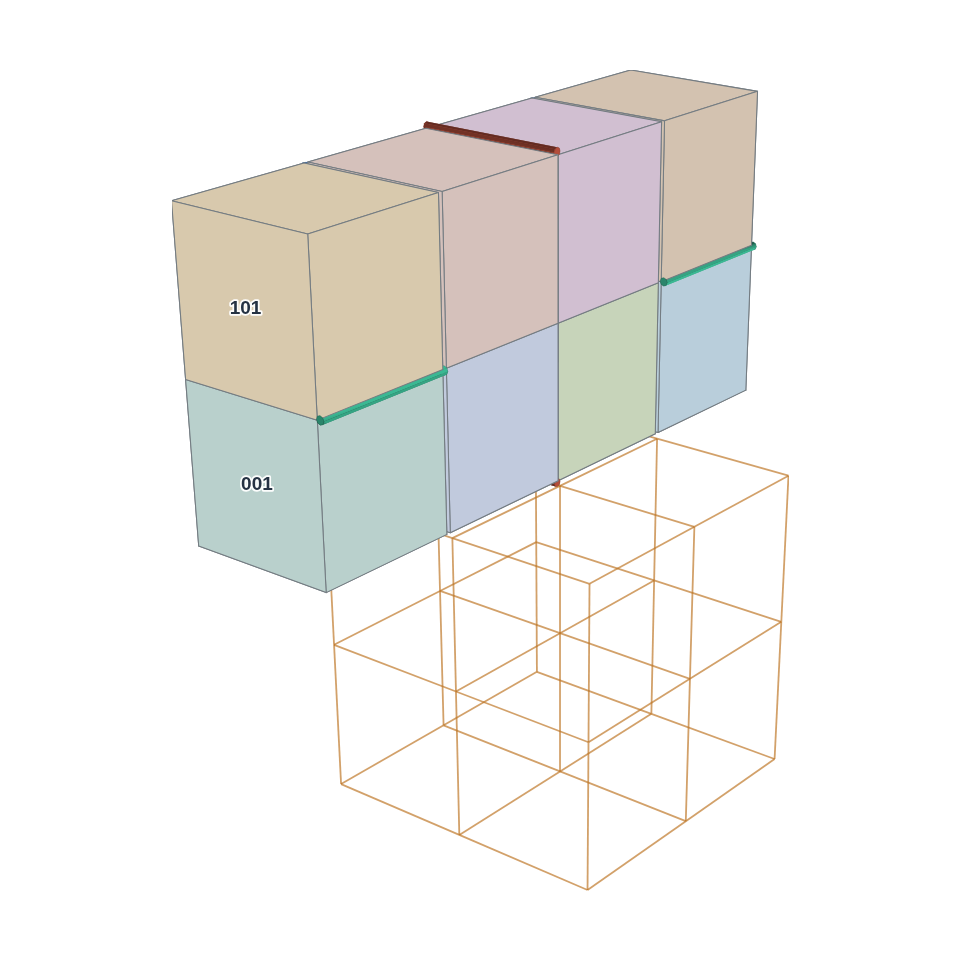}\caption{$S_{RG}$}\end{subfigure}
\begin{subfigure}{.24\textwidth}\includegraphics[width=\linewidth]{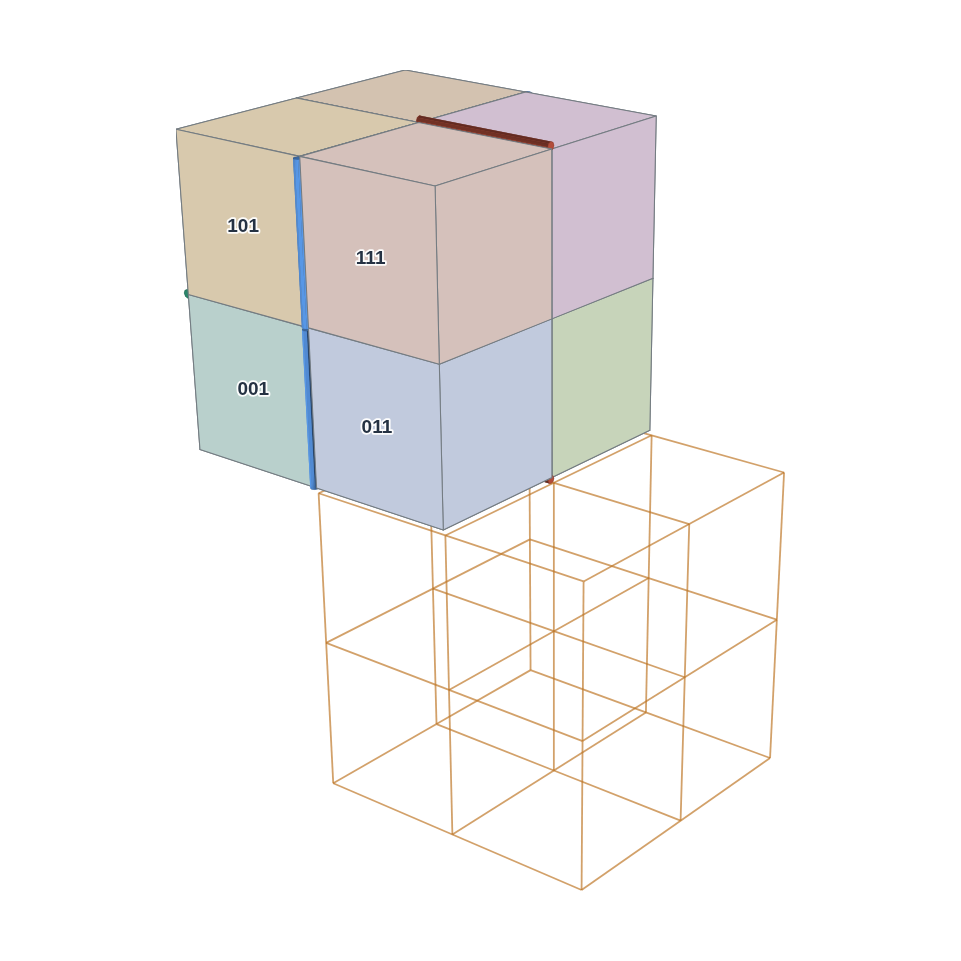}\caption{$S_{RGB}$}\end{subfigure}
\caption{Red--green--blue route. The amber lattice is fixed in world coordinates, making the displacement of the final compact assembly visible. All panels use the same viewing direction and perspective model, with automatic recentering and framing to keep the entire mechanism visible.}\label{fig:rgb}
\end{figure*}

The first route is
\begin{equation}\label{eq:rgb}
S_0\xrightarrow{R(-90^\circ)}S_R
\xrightarrow{G(-180^\circ)}S_{RG}
\xrightarrow{B(-180^\circ)}S_{RGB}.
\end{equation}
The first red action opens two four-block slabs. The green stage is applied about axes transported through the completed red action. For the final blue stage, the user-verified rigid groups are particularly simple:
\begin{align*}
\text{stationary base: }&\{110,111,010,011\},\\
\text{wing A: }&\{100,000\},\\
\text{wing B: }&\{101,001\}.
\end{align*}
Each two-block wing rotates about its own transported blue hinge; the wings must fold outward rather than through the stationary base. Figure~\ref{fig:rgb} shows the four sampled endpoint poses of this route.

\subsection{Route B: blue--red--green}
\begin{figure*}[t]
\centering
\begin{subfigure}{.24\textwidth}\includegraphics[width=\linewidth]{figures/initial.png}\caption{$S_0$}\end{subfigure}
\begin{subfigure}{.24\textwidth}\includegraphics[width=\linewidth]{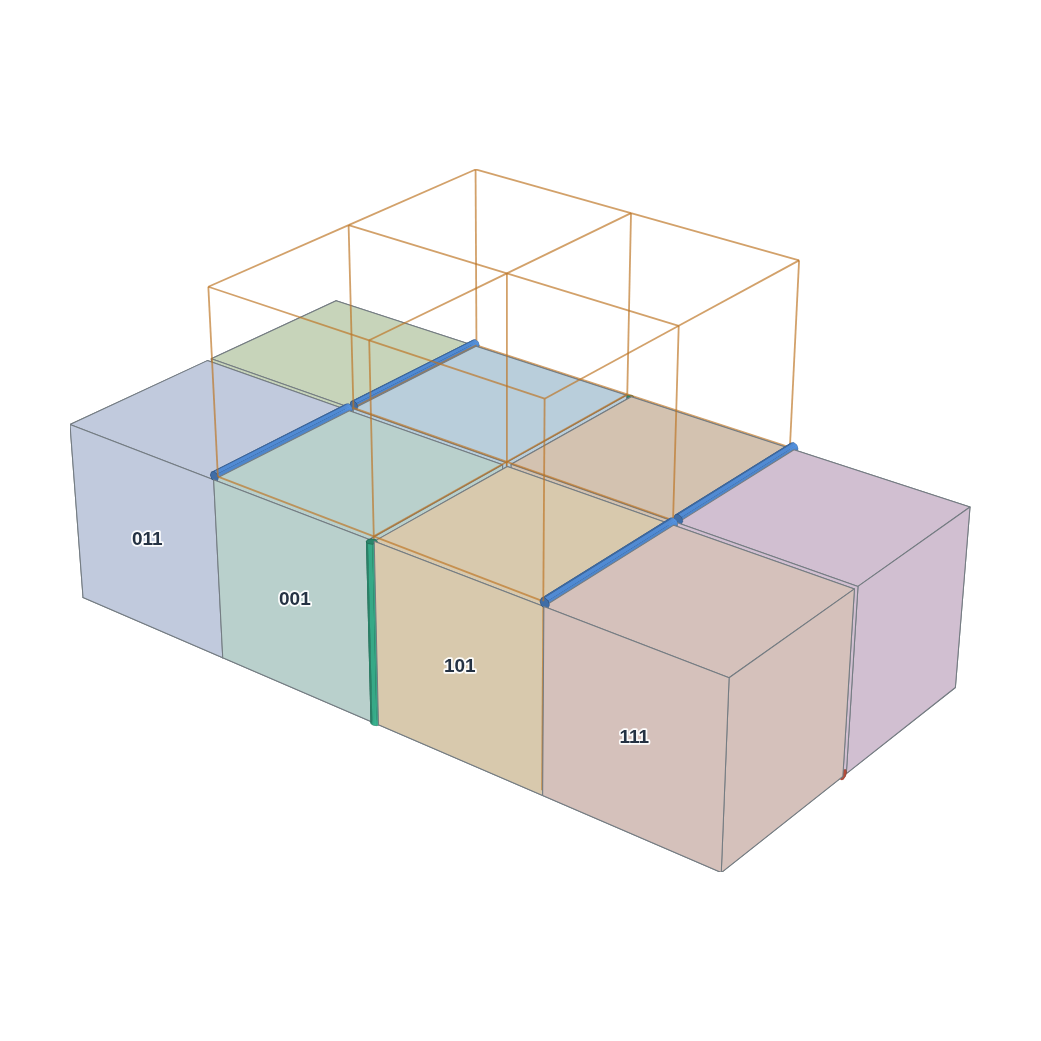}\caption{$S_B$}\end{subfigure}
\begin{subfigure}{.24\textwidth}\includegraphics[width=\linewidth]{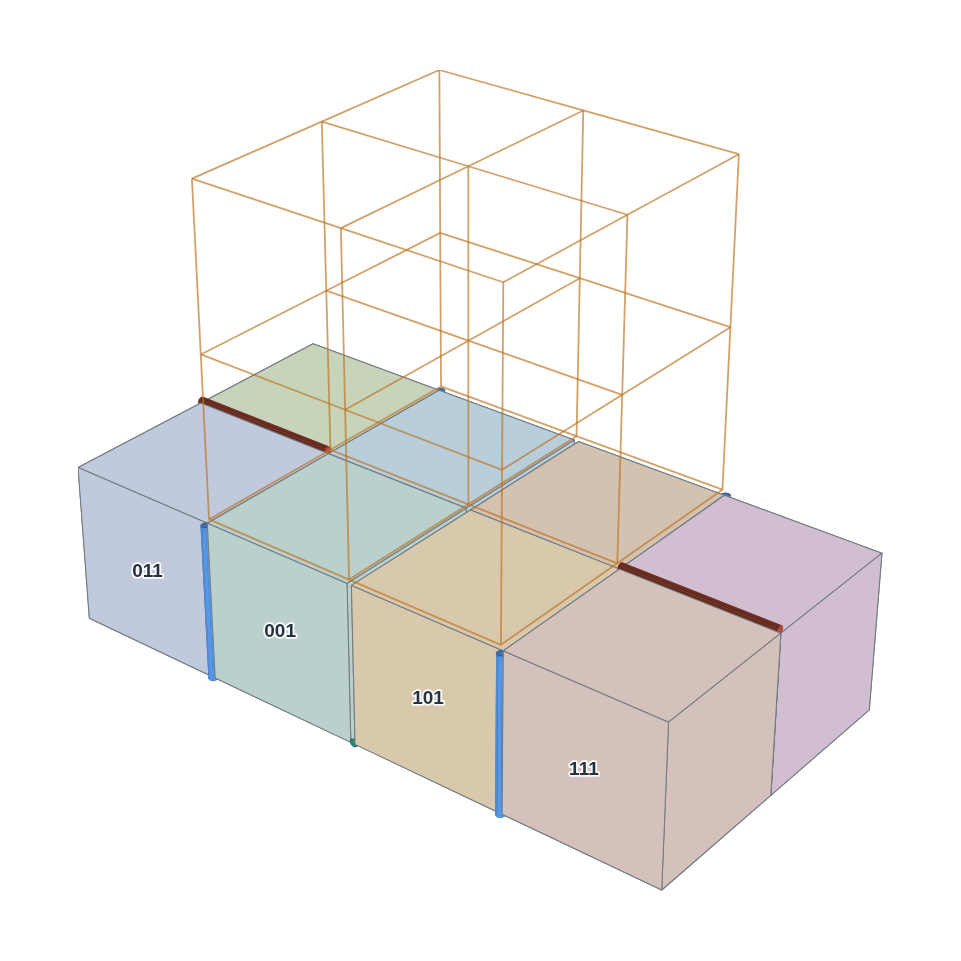}\caption{$S_{BR}$}\end{subfigure}
\begin{subfigure}{.24\textwidth}\includegraphics[width=\linewidth]{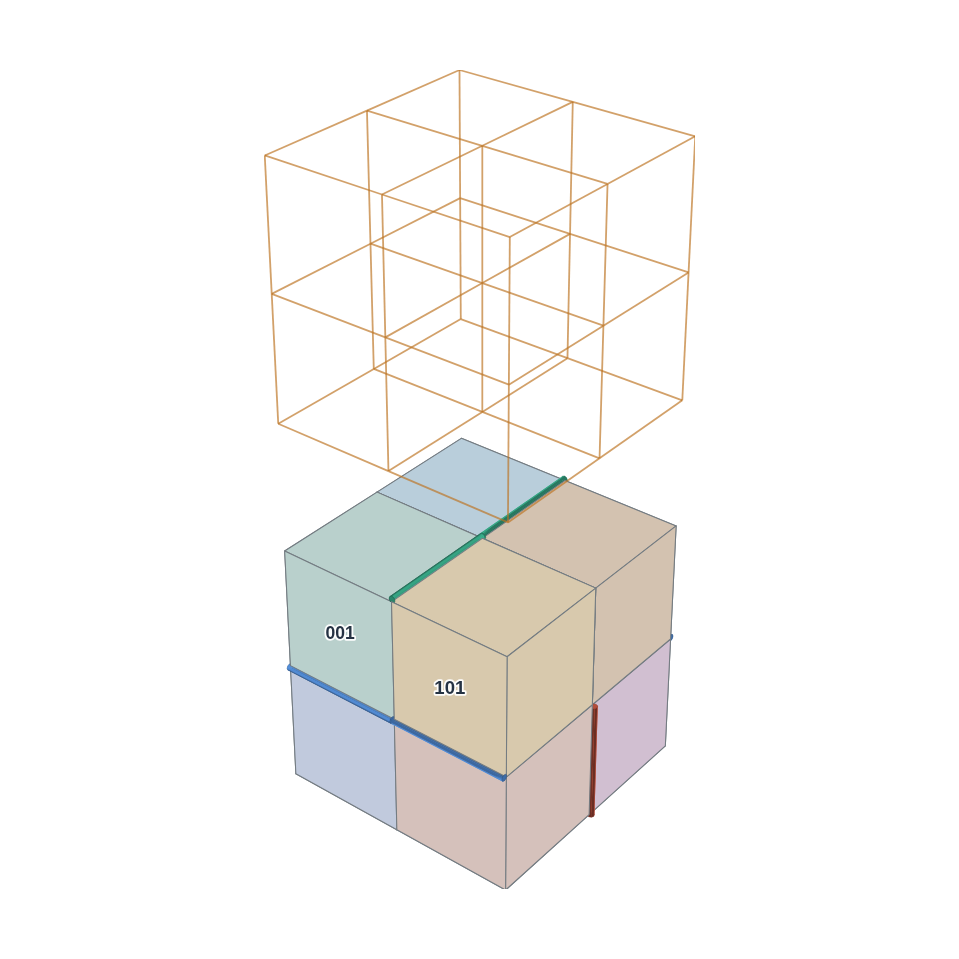}\caption{$S_{BRG}$}\end{subfigure}
\caption{Blue--red--green route. The final compact assembly lies below the stationary reference cube, unlike the endpoint of Fig.~\ref{fig:rgb}. This world-space displacement does not establish non-congruence of the labeled terminal configurations.}\label{fig:brg}
\end{figure*}

The second route is
\begin{equation}\label{eq:brg}
S_0\xrightarrow{B(-180^\circ)}S_B
\xrightarrow{R(-90^\circ)}S_{BR}
\xrightarrow{G(90^\circ)}S_{BRG}.
\end{equation}
The second-stage red action acts on two rigid four-block groups:
\begin{align*}
Z_1&=\{011,001,101,111\},\\
Z_0&=\{110,100,000,010\}.
\end{align*}
These are the two original $z$ layers. In the final green action, both four-block groups move, with opposite signed hinge angles:
\begin{align*}
X_1&=\{101,111,100,110\}:\quad +90^\circ,\\
X_0&=\{011,001,010,000\}:\quad -90^\circ.
\end{align*}
Treating $X_0$ as stationary would not reproduce the observed two-sided closing action. The two four-block groups instead rotate like two leaves of a book about their respective transported green hinge axes. Figure~\ref{fig:brg} records the sampled states.

\section{Terminal configurations and equivalence}
\subsection{Position alone is insufficient}
Let $\mathcal C_A$ and $\mathcal C_B$ denote the final configurations of Eqs.~\eqref{eq:rgb} and~\eqref{eq:brg}. The two compact endpoints occur in different parts of the fixed world coordinate system. Moreover, the labels appear on differently oriented visible faces. Neither observation proves that the terminal configurations differ intrinsically.

A precise equivalence test seeks one $R\in SO(3)$ and one $\mathbf t\in\mathbb R^3$ such that, for every persistent label $i$,
\begin{equation}\label{eq:equiv}
\mathbf c_i^B=R\mathbf c_i^A+\mathbf t,
\qquad Q_i^B=RQ_i^A.
\end{equation}
The \emph{same} rotation must satisfy both equations for all eight blocks. A reflection (determinant $-1$) is not allowed because it is not a physical rigid rotation.

\subsection{Numerical result for the present implementation}
We exported all eight final centers and three local basis directions from the actual JavaScript transformation functions. A rigid alignment of corresponding labeled centers yields the proper quarter-turn
\begin{equation}\label{eq:R}
R=\begin{bmatrix}0&1&0\\-1&0&0\\0&0&1\end{bmatrix},
\end{equation}
with an appropriate translation $\mathbf t=\overline{\mathbf c}^{B}-R\overline{\mathbf c}^{A}$. The rotation is a $-90^\circ$ turn about the world $z$ axis, with column-vector convention. After alignment, the center RMS residual is below $10^{-12}$ in model units. More importantly, the eight orientation frames also agree to floating-point precision (maximum matrix-entry discrepancy below $10^{-12}$).

\paragraph{Interpretation.} In this implemented kinematic model, the two terminal assemblies are therefore \emph{rigidly equivalent as labeled, oriented assemblies}, even though they have different world-space positions and different orientations before alignment. This is stronger than matching their $2\times2\times2$ outlines. It also corrects a tempting visual inference that the endpoints necessarily represent different internal arrangements. Whether the real manufactured object has additional constraints not captured by this model remains to be tested.

\begin{figure}[t]
\centering
\includegraphics[width=.49\linewidth]{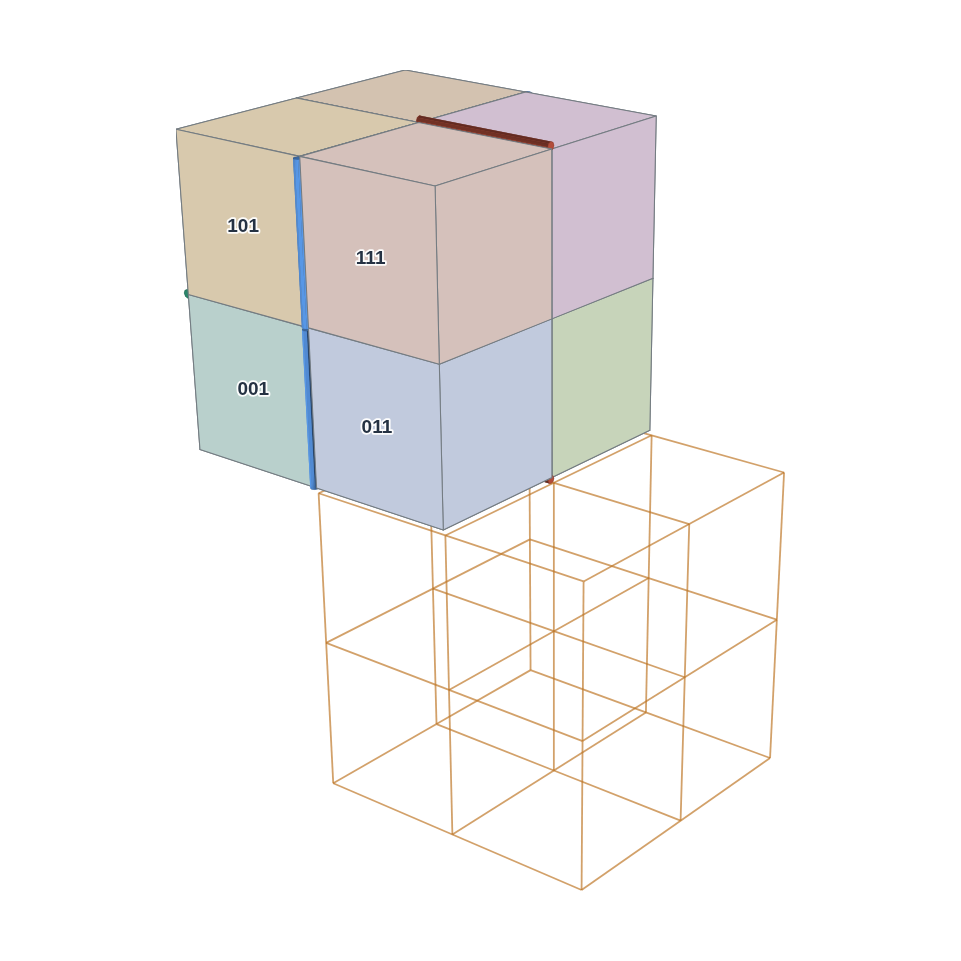}\hfill
\includegraphics[width=.49\linewidth]{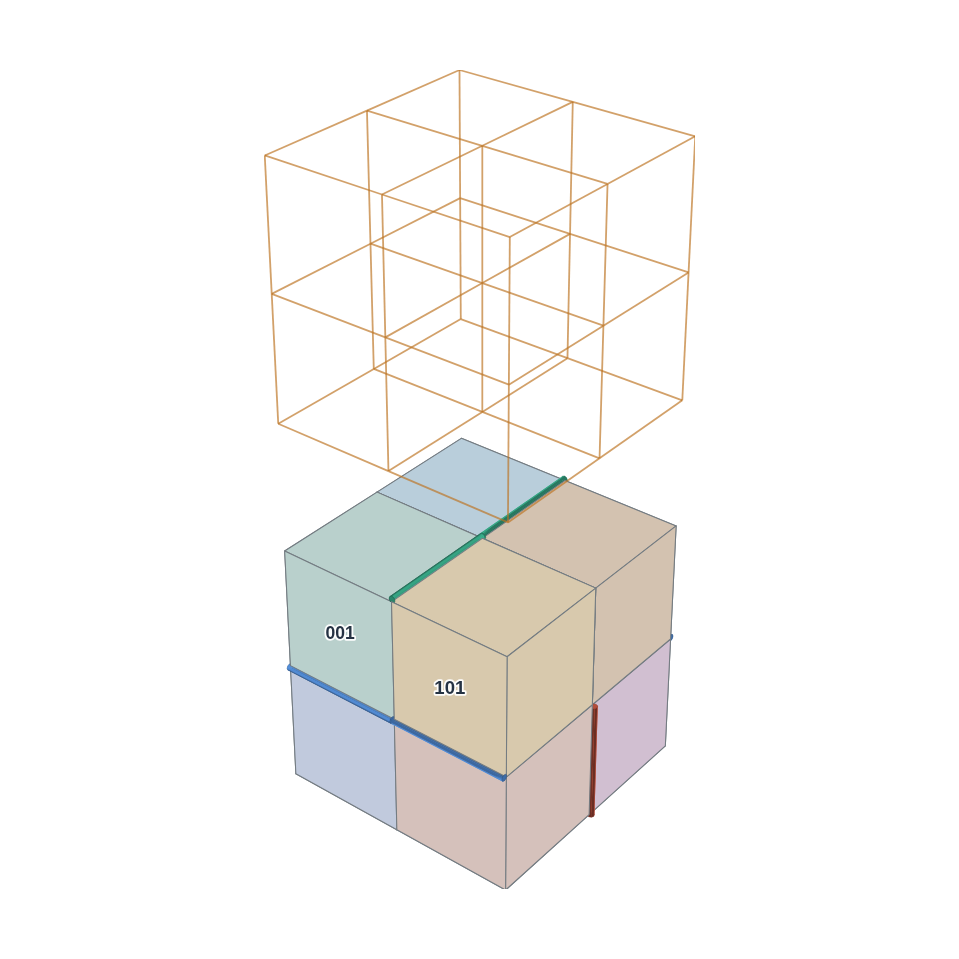}
\caption{Terminal states in a common, fixed camera frame: $S_{RGB}$ (left) and $S_{BRG}$ (right). Their different placement and apparent face orientations are reconciled by the single rigid transformation of Eq.~\eqref{eq:equiv}.}\label{fig:compare}
\end{figure}

\subsection{A finite rotation test}
For an exact lattice-aligned final cube, one may avoid unconstrained optimization by enumerating the 24 proper signed permutation matrices
\begin{equation}
\mathcal O=\{R\in\mathbb Z^{3\times3}:R^TR=I,\ \det R=1\}.
\end{equation}
For each candidate, align centroids and test the eight centers and orientation frames. This distinguishes three cases: unmatched centers, matched centers but unmatched frames, and full pose equivalence. Our numerical export falls in the third category.

\section{Configuration paths and surface exposure}
The two verified routes form the observed subgraph
\begin{equation*}
\begin{array}{ccccccc}
&&S_R&\xrightarrow{G}&S_{RG}&\xrightarrow{B}&S_{RGB}\\[-2pt]
&\nearrow^{R}&&&&&\\[-2pt]
S_0&&&&&&\\[-2pt]
&\searrow_{B}&&&&&\\[-2pt]
&&S_B&\xrightarrow{R}&S_{BR}&\xrightarrow{G}&S_{BRG}.
\end{array}
\end{equation*}
This is a graph of \emph{demonstrated stage transitions}, not the full configuration space. A rigorous global classification would require determining all feasible hinge motions, compatibility constraints, and self-collision restrictions. The simulation currently implements selected rigid group rotations and should not be mistaken for a general constraint solver.

The magic photo cube also involves a surface-level transformation. Each material cube has six local faces with outward normals $\pm\mathbf e_x,\pm\mathbf e_y,\pm\mathbf e_z$. At pose $Q_i$, a local face normal becomes $Q_i\mathbf n$. Thus, a complete visual analysis should track both which material faces lie on the exterior boundary and how their orientations change. This is the mechanism behind the emergence of a previously hidden photograph. A companion gallery implementation assigns twelve square images to the two principal configurations by preserving texture coordinates on material faces. The full face-exposure map and its physical validation remain implementation-dependent.

\begin{figure*}[htbp!]
\centering
\begin{subfigure}[t]{0.235\textwidth}\centering\includegraphics[width=\linewidth]{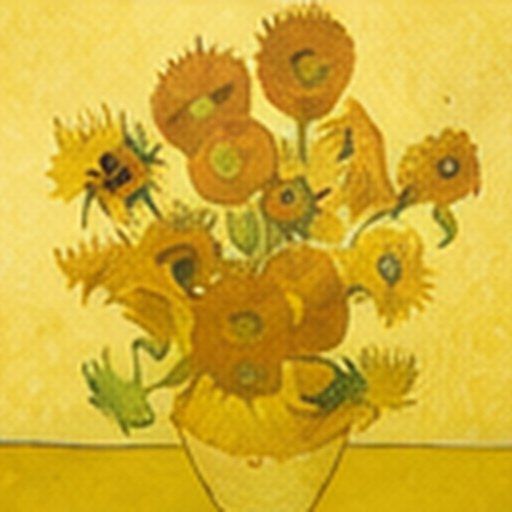}\caption{\texttt{000}}\end{subfigure}\hfill
\begin{subfigure}[t]{0.235\textwidth}\centering\includegraphics[width=\linewidth]{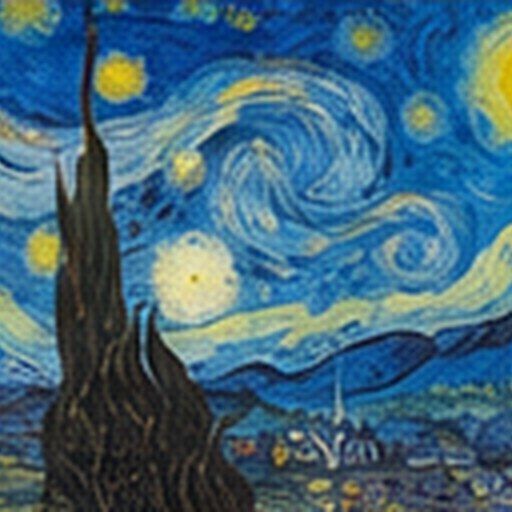}\caption{\texttt{001}}\end{subfigure}\hfill
\begin{subfigure}[t]{0.235\textwidth}\centering\includegraphics[width=\linewidth]{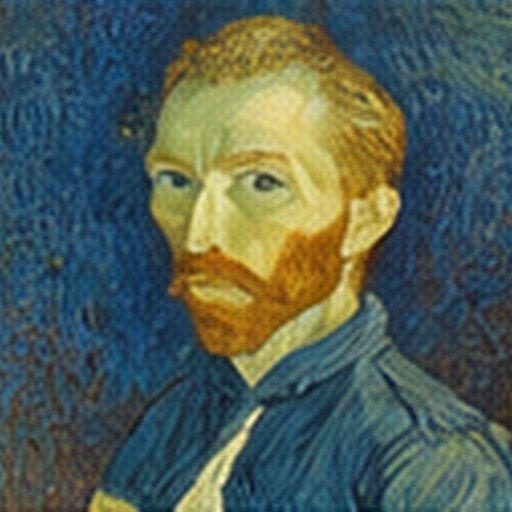}\caption{\texttt{010}}\end{subfigure}\hfill
\begin{subfigure}[t]{0.235\textwidth}\centering\includegraphics[width=\linewidth]{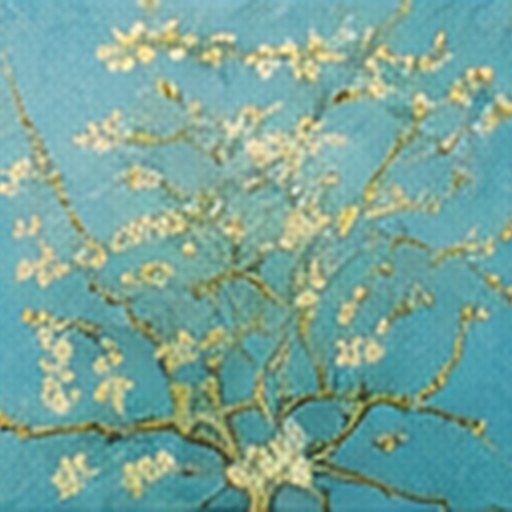}\caption{\texttt{011}}\end{subfigure}
\par\medskip
\begin{subfigure}[t]{0.235\textwidth}\centering\includegraphics[width=\linewidth]{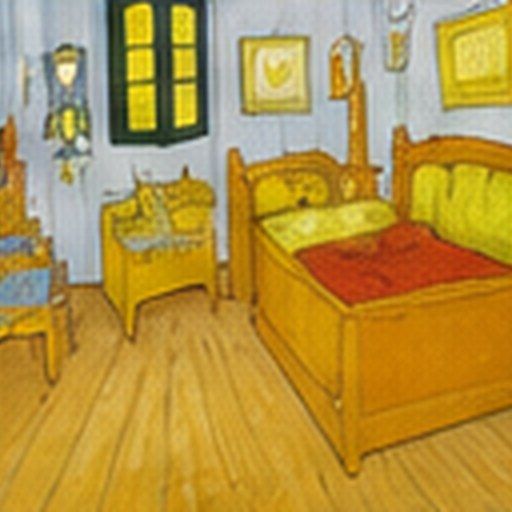}\caption{\texttt{100}}\end{subfigure}\hfill
\begin{subfigure}[t]{0.235\textwidth}\centering\includegraphics[width=\linewidth]{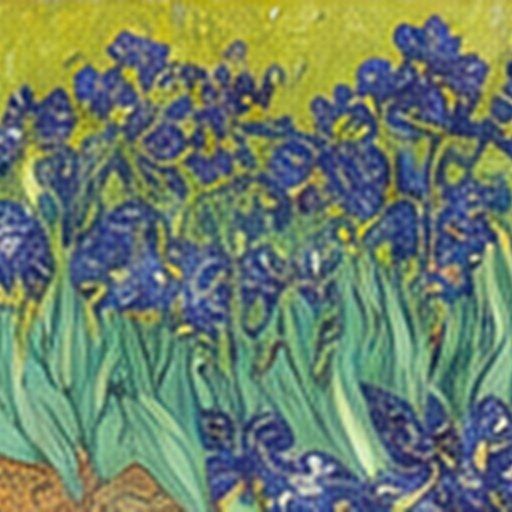}\caption{\texttt{101}}\end{subfigure}\hfill
\begin{subfigure}[t]{0.235\textwidth}\centering\includegraphics[width=\linewidth]{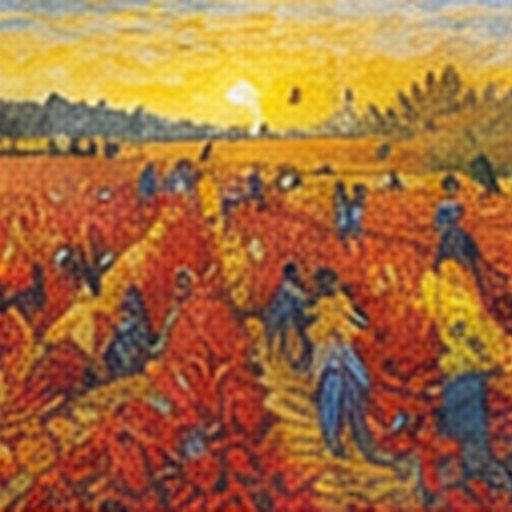}\caption{\texttt{110}}\end{subfigure}\hfill
\begin{subfigure}[t]{0.235\textwidth}\centering\includegraphics[width=\linewidth]{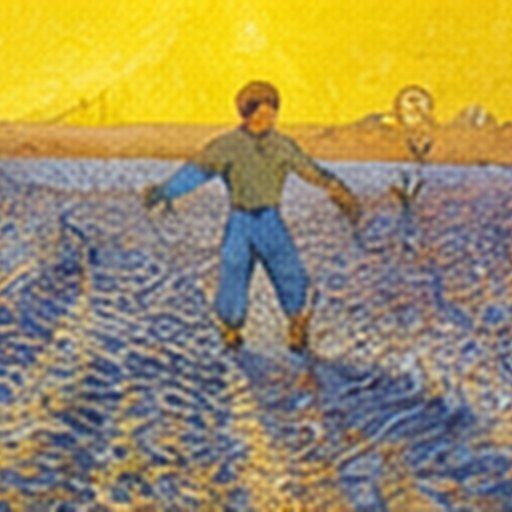}\caption{\texttt{111}}\end{subfigure}
\par\medskip
\begin{subfigure}[t]{0.235\textwidth}\centering\includegraphics[width=\linewidth]{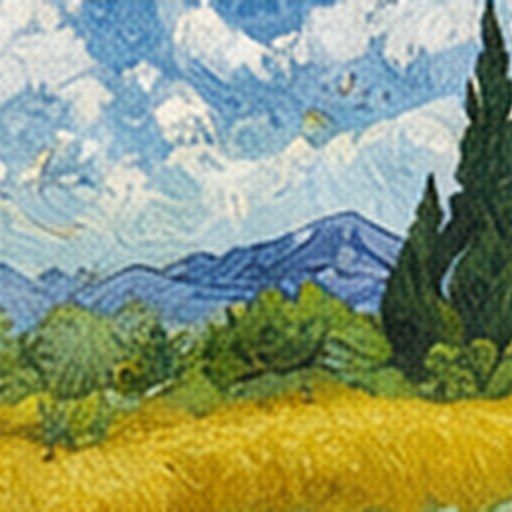}\caption{\texttt{200}}\end{subfigure}\hfill
\begin{subfigure}[t]{0.235\textwidth}\centering\includegraphics[width=\linewidth]{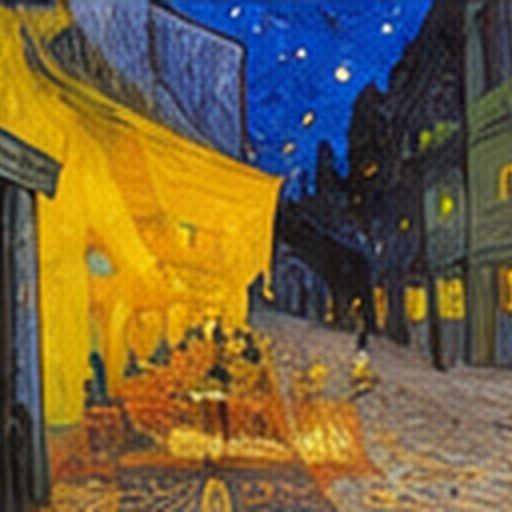}\caption{\texttt{201}}\end{subfigure}\hfill
\begin{subfigure}[t]{0.235\textwidth}\centering\includegraphics[width=\linewidth]{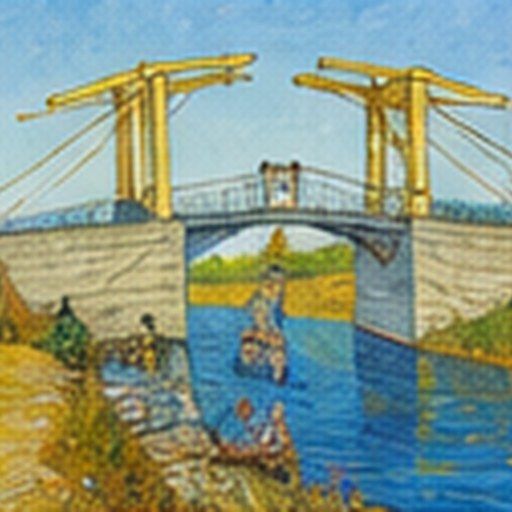}\caption{\texttt{210}}\end{subfigure}\hfill
\begin{subfigure}[t]{0.235\textwidth}\centering\includegraphics[width=\linewidth]{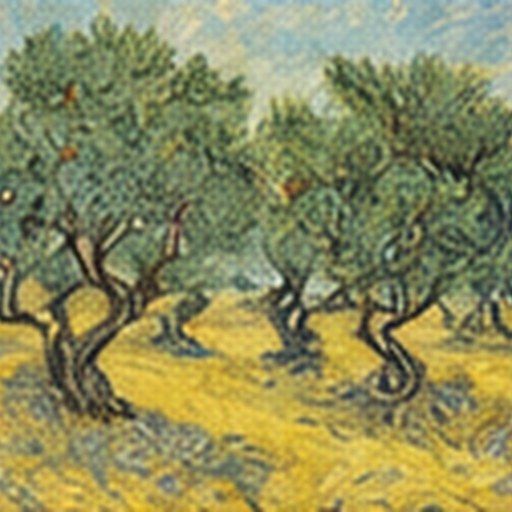}\caption{\texttt{211}}\end{subfigure}
\par\medskip
\caption{Twelve separate Van Gogh paintings assigned to the twelve $2\times2$ image frames. Subcaptions are the geometric identifiers: first digit $0,1,2$ denotes the axis; second digit $0,1$ denotes outside/inside; third digit $0,1$ denotes the negative/positive normal direction. The order in this atlas is by identifier, not a sequential comic narrative.}
\label{fig:akleman-atlas}
\end{figure*}

\section{Persistent image assignment and the twelve-image index}
\label{sec:images}
A magic photo cube offers twelve $2\times2$ image frames: six visible in one compact display configuration and six in the complementary configuration. Each frame comprises four square patches belonging to physical cube faces. A single face-assignment procedure is run on each of the two display configurations, and the resulting UV coordinates remain attached to the material faces during subsequent folding. Reassigning textures dynamically during motion would incorrectly change the physical artwork.

We identify an image with a three-digit code $a b c$, where $a\in\{0,1,2\}$ specifies the normal axis ($x,y,z$), $b\in\{0,1\}$ specifies outside or inside display, and $c\in\{0,1\}$ specifies negative or positive normal direction. The twelve identifiers are $000,001,010,011,100,101,110,111,200,201,210,211$. These are \emph{face identifiers}, not the reading order of a story. The VanGogh, Akleman, and Diagnostic collections use the same filenames and geometry-to-texture correspondence; the diagnostic set contains asymmetric labels to test orientation.

\paragraph{Independent image collection.}
Figure~\ref{fig:akleman-atlas} presents the twelve Van Gogh paintings as individual images. The three-digit labels beneath each painting denote the corresponding geometric $2\times2$ frame, rather than a required narrative order. Each painting is a separate image file and is composed using the \texttt{subcaption} package.

\section{Implementation and reproducibility}
The accompanying dependency-free HTML/JavaScript application uses a canvas-based perspective renderer, persistent binary labels, local-edge hinge geometry, and Rodrigues rotations. Each later-stage axis is reconstructed from the preceding frozen configuration. The display offers a fixed $2\times2\times2$ grid, an optional larger $6\times6\times6$ grid, camera orbit, right-button panning, and zoom. Depth sorting allows the reference grid to be occluded by moving cube faces. The figures in this manuscript were captured automatically from the simulation at exact slider endpoints with a common camera pose. The numerical file \texttt{poses.json} contains all terminal centers and orientation frames used in the equivalence test. 

The accompanying project website~\cite{Akleman2026MagicCubeWebsite} provides two interactive implementations: a geometric simulator that visualizes the labeled cubical cells, hinge connections, and folding sequences, and a textured simulator that demonstrates the assignment of twelve images to the physical faces of the cube. These implementations allow readers to interactively explore the geometric constructions and folding operations presented in this paper.

The implementation is a visually validated geometric reconstruction. Hinge clearances, material thickness, real adhesive or mechanical hinge limits, and collisions during continuous motion have not been rigorously measured. A numerical match of endpoint poses does not establish collision-free physical realizability for arbitrary interpolations.

\section{Discussion and Future Work}
\label{sec:discussion}

The principal contribution of this work is a representation of piano-hinged solids as edge-connected non-manifold cell complexes. Rather than treating hinges as independent geometric objects, the proposed approach represents their connections through explicit edge identifications between the boundary meshes of rigid solid components. This provides a unified description of the constituent solids, their persistent connectivity, and their changing geometric configurations. The magic photo cube serves as a representative example, demonstrating how a relatively simple assembly can exhibit complex articulated behavior while maintaining a fixed underlying incidence structure.

\subsection{Non-Manifold Representation of Articulated Solids}

The proposed representation separates the topology of an articulated assembly from its geometric realization. Each component is a rigid solid bounded by a two-manifold surface, while the complete structure is constructed by identifying selected edges of different components. The resulting non-manifold complex preserves the identities of the original solids and the precise locations of their connections. During articulation, these incidence relationships remain unchanged, even though the positions and orientations of the components vary.

An important advantage of this construction is that hinge locations are intrinsic to the representation. An ordinary connectivity graph identifies which components are connected, but does not specify the particular edges participating in those connections. In contrast, the non-manifold complex records the complete edge-incidence structure, including the correspondence between the endpoints of identified edges. This distinction is essential because different edge identifications can produce articulated assemblies with different geometric constraints and folding capabilities, even when their connectivity graphs are isomorphic.

The representation is not restricted to cubical components. In principle, any collection of rigid polyhedral solids with compatible edges can be connected through the same construction. This suggests a general approach to modeling articulated structures whose components may have substantially different shapes, sizes, and local geometries. The present work establishes the representation and demonstrates it using an eight-cube assembly; the systematic construction and analysis of more general articulated polyhedral complexes remain subjects for future investigation.

\subsection{Connectivity, Constraints, and Articulated Motion}

The classification of connectivity graphs provides a useful organizational framework for understanding articulated structures. Open chains and trees contain no connectivity cycles, allowing relative transformations to be propagated from a selected root component. Closed cycles and more general graphs introduce loop-closure conditions that couple the transformations of their constituent solids. The number of independent graph cycles characterizes the combinatorial organization of these constraints, but does not determine the actual mobility of the corresponding mechanism.

The magic photo cube illustrates this distinction. Its eight rigid cubical cells are connected through eight piano hinges, producing a cyclic connectivity graph isomorphic to $C_8$. Although the graph contains only one independent cycle, the available folding motions depend on the geometric placement of the hinge edges and the resulting compatibility constraints. The observed three-stage folding sequences therefore involve coordinated transformations of rigid groups rather than arbitrary independent rotations of individual hinges.

The two reconstructed folding routes also demonstrate the importance of distinguishing configurations in a fixed reference frame from their intrinsic geometric relationships. Although the routes produce compact assemblies at different world-space positions and orientations, the terminal configurations are equivalent under a single proper rigid transformation. This equivalence holds for the complete set of labeled component poses, including both positions and orientations, rather than merely for the external cubical silhouette.

Such comparisons suggest that the configuration space of an articulated non-manifold complex should be considered modulo global rigid transformations when the objective is to distinguish internal configurations. This viewpoint provides a natural foundation for identifying equivalent states, analyzing folding sequences, and investigating the structure of admissible motions.

\subsection{Limitations of the Present Implementation}

The current implementation demonstrates a geometric representation and selected articulated transformations, rather than a general-purpose solver for arbitrary edge-connected non-manifold complexes. The folding operations of the magic photo cube are implemented through prescribed rigid-group rotations about transported hinge axes. Their compatibility has been examined for the reconstructed sequences, but the implementation does not automatically determine every admissible motion from an arbitrary set of edge identifications.

Furthermore, the numerical equivalence of the two terminal configurations does not establish that all intermediate configurations are physically realizable. The present geometric model does not rigorously account for the finite dimensions of manufactured hinges, mechanical clearances, material deformation, or collisions between solid components throughout continuous folding motions. These effects may restrict the motions available to a physical realization.

The present analysis also does not establish a complete classification of the magic photo cube's configuration space. The two reconstructed routes constitute demonstrated paths through selected configurations, but additional folding routes, intermediate states, or motion constraints may exist. A complete analysis would require systematic exploration of the constraint equations together with geometric collision detection.

\subsection{Future Research Directions}

A natural next step is to develop a general computational framework that accepts an arbitrary collection of manifold polyhedral meshes together with prescribed edge identifications and constructs the corresponding articulated non-manifold complex. Such a system could automatically generate hinge constraints, identify connected rigid groups, and compute admissible motions while preserving the incidence relationships of the original components.

A second direction concerns the analysis of closed-loop mechanisms. For structures containing one or more independent connectivity cycles, the associated geometric constraints must remain satisfied throughout articulation. Developing numerical methods for solving these constraints, identifying independent motion parameters, and detecting singular configurations would extend the present representation into a more general animation and mechanism-analysis framework.

Collision-aware motion planning is another important direction. Incorporating the actual volumes of the component solids, finite hinge clearances, and nonpenetration constraints would make it possible to distinguish geometrically compatible configurations from physically realizable folding paths. Such methods would also support the systematic exploration of reachable configurations and the identification of alternative folding sequences.

The framework also suggests opportunities for constructing articulated structures from more general polyhedral cells. For example, Voronoi or Wigner--Seitz cells could provide families of polyhedral components with different geometric and combinatorial properties. Connecting selected edges of these cells could produce articulated assemblies whose folding behavior differs substantially from that of conventional cubical or rectangular mechanisms. The existence, mobility, and physical realizability of such structures remain open questions.

Finally, the magic photo cube illustrates a relationship between articulated geometry and visual communication. Because images are attached to persistent material faces, folding changes which portions of the artwork are exposed without changing their underlying texture coordinates. Extending the implementation to compute material-face visibility automatically could support the design of folding displays, sequential visual narratives, and other applications in which geometric transformations control the presentation of visual information.

These directions suggest that edge-connected non-manifold cell complexes may provide a useful foundation for the representation, design, and animation of articulated polyhedral structures beyond the particular mechanism examined in this paper.

\section{Conclusion}
\label{sec:conclusion}

We have presented a framework for representing and animating piano-hinged solids as edge-connected non-manifold cell complexes. The central idea is to construct articulated structures by identifying selected edges of otherwise manifold boundary meshes. These identifications encode the persistent connections between rigid components without requiring hinges to be modeled as separate geometric objects. The resulting representation preserves the identities and incidence relationships of the constituent solids while allowing their positions and orientations to change through articulated motion.

We have distinguished the complete edge-incidence structure from its associated connectivity graph and classified articulated assemblies according to open chains, trees, closed cycles, and general graphs. This distinction clarifies the relationship between the combinatorial organization of an assembly and the geometric constraints governing its motion. In particular, closed-loop structures require compatible transformations around their connectivity cycles, and their mobility depends on the geometric arrangement of their hinge axes rather than on graph topology alone.

Using the magic photo cube as a representative example, we have reconstructed two distinct three-stage folding sequences of eight rigid cubical cells connected through eight piano hinges. By tracking persistent component identities and local coordinate frames, we demonstrated that the two terminal configurations, although differently positioned and oriented in a fixed reference frame, are equivalent under a single proper rigid transformation. This result illustrates the importance of comparing complete labeled poses rather than external shapes or component positions alone.

The accompanying interactive implementation demonstrates how the proposed representation can support geometric visualization, articulated animation, configuration comparison, and persistent texture mapping. Although the present study does not provide a general constraint solver or a complete classification of admissible folding motions, it establishes a consistent geometric and topological framework for investigating edge-connected articulated solids.

More broadly, the work suggests that non-manifold cell complexes can serve not only as representations of static geometric objects, but also as representations of articulated structures whose connectivity remains fixed while their geometric realization changes. Extending this approach to general polyhedral components, more complex connectivity graphs, and automatically computed admissible motions offers a promising direction for future research in geometric modeling and animation.


\paragraph{Code Availability.}
The interactive Magic Photo Cube implementations, including the geometric and textured simulators, are publicly accessible through the accompanying project website~\cite{Akleman2026MagicCubeWebsite}.

\paragraph{AI Assistance Disclosure.}
OpenAI's ChatGPT was used as an interactive assistant during
the development of the JavaScript and WebGL implementation
and the preparation of this manuscript. Its contributions
included generating and revising source code, assisting with
debugging and interface development, preparing LaTeX text,
and improving the organization and clarity of the manuscript.
The geometric concepts, mathematical formulation, modeling
decisions, and interpretation of the results were developed
and directed by the authors through an iterative process
of implementation, testing, and revision. All AI-generated
material was reviewed and revised by the author, who take
full responsibility for the scientific content, correctness,
and conclusions of this work.

\bibliographystyle{unsrtnat}
\bibliography{references}

@misc{chica_video,
  author       = {{Chica and Jo}},
  title        = {Magic Folding Photo Cube Tutorial},
  howpublished = {YouTube video},
  url          = {https://www.youtube.com/watch?v=lmwBLHAqbWg}
}

@misc{tanu_video,
  author       = {{Tanu Creative}},
  title        = {Magic Photo Cube Tutorial},
  howpublished = {YouTube video},
  url          = {https://www.youtube.com/watch?v=WAuQvAZKOOM}
}

@misc{vangogh_cube,
  author = {{Museum Webshop}},
  title  = {Magic Cube, {Van Gogh}},
  url    = {https://www.museum-webshop.com/en/magic-cube-van-gogh.html}
}

@techreport{vgm_2003,
  author      = {{Van Gogh Museum}},
  title       = {Annual Report 2003},
  institution = {Van Gogh Museum},
  year        = {2003},
  url         = {https://www.vangoghmuseum.nl/assets/d4b8895d-a9ae-436a-9fa9-d634cab9505f/Annual-Report-2003_Van-Gogh-Museum}
}

@book{frederickson1997,
  author    = {Greg N. Frederickson},
  title     = {Dissections: Plane and Fancy},
  publisher = {Cambridge University Press},
  year      = {1997}
}

@book{frederickson2002,
  author    = {Greg N. Frederickson},
  title     = {Hinged Dissections: Swinging and Twisting},
  publisher = {Cambridge University Press},
  year      = {2002}
}

@book{frederickson2006,
  author    = {Greg N. Frederickson},
  title     = {Piano-Hinged Dissections: Time to Fold!},
  publisher = {A K Peters},
  year      = {2006}
}

@article{demaine2005polyominoes,
  author  = {Erik D. Demaine and Martin L. Demaine and David Eppstein and Greg N. Frederickson and Erich Friedman},
  title   = {Hinged Dissection of Polyominoes and Polyforms},
  journal = {Computational Geometry},
  year    = {2005},
  volume  = {31},
  number  = {3},
  pages   = {237--262},
}

@article{abbott2012hinged,
  author  = {Timothy G. Abbott and Zachary Abel and David Charlton and Erik D. Demaine and Martin L. Demaine and Scott Duke Kominers},
  title   = {Hinged Dissections Exist},
  journal = {Discrete \& Computational Geometry},
  year    = {2012},
  volume  = {47},
  pages   = {150--186},
  note    = {Verify pagination and author list against publisher before final submission}
}

@article{Trenkler2000MagicCubes,
  author  = {Trenkler, Mari{\'a}n},
  title   = {A Construction of Magic Cubes},
  journal = {The Mathematical Gazette},
  volume  = {84},
  number  = {499},
  pages   = {36--41},
  year    = {2000},
  doi     = {10.2307/3621472}
}

@article{Rani2022FusedMagicCube,
  author  = {Rani, Narbda and
             Sharma, Suvita Rani and
             Mishra, Vinod},
  title   = {Grayscale and Colored Image
             Encryption Model Using a
             Novel Fused Magic Cube},
  journal = {Nonlinear Dynamics},
  volume  = {108},
  pages   = {1773--1796},
  year    = {2022},
  doi     = {10.1007/s11071-022-07276-y}
}

@article{Chu2025MagicCubeMetamaterials,
  author  = {Chu, Zuntian and
             Cai, Xinqi and
             Zhu, Ruichao and
             Liu, Tonghao and
             Sun, Huiting and
             Wu, Fan and
             Li, Tiefu and
             Jia, Yuxiang and
             Han, Yajuan and
             Qu, Shaobo and
             Cui, Tiejun and
             Wang, Jiafu},
  title   = {Magic Cube Metamaterials},
  journal = {Science Bulletin},
  year    = {2025},
  doi     = {10.1016/j.scib.2025.07.010}
}

@article{schneider2025topology,
  title={Topology-engineered piezoresistive lattices with programmable strain sensing, auxeticity, and failure modes},
  author={Schneider, Johannes and Utzeri, Mattia and Krishnamurthy, Vinayak Raman and Akleman, Ergun and Kumar, S},
  journal={Materials Horizons},
  volume={12},
  number={19},
  pages={8012--8032},
  year={2025},
  publisher={The Royal Society of Chemistry}
}

@article{ebert2024abc,
  title={ABC-Auxetics: An Implicit Design Approach for Negative Poisson's Ratio Materials},
  author={Ebert, Matthew and Adhikari, Riddhi and Hasan, Md Kamrul and Lupo, Kai and Akleman, Ergun and Pharr, Matt and Krishnamurthy, Vinayak R},
  journal={Advanced Engineering Materials},
  volume={26},
  number={8},
  pages={2301359},
  year={2024},
  publisher={Wiley Online Library}
}

@article{schneider2024concurrent,
  title={Concurrent geometrico-topological tuning of nanoengineered auxetic lattices fabricated by material extrusion for enhancing multifunctionality: Multiscale experiments, finite element modeling and data-driven prediction},
  author={Schneider, Johannes and Ebert, Matthew and Tipireddy, Ramakrishna and Krishnamurthy, VR and Akleman, Ergun and Kumar, Shanmugam},
  journal={Additive Manufacturing},
  volume={88},
  pages={104213},
  year={2024},
  publisher={Elsevier}
}

@article{GrimaEvans2000,
  author  = {Grima, Joseph N. and Evans, Kenneth E.},
  title   = {Auxetic Behavior from Rotating Squares},
  journal = {Journal of Materials Science Letters},
  volume  = {19},
  number  = {17},
  pages   = {1563--1565},
  year    = {2000},
  doi     = {10.1023/A:1006781224002}
}

@misc{Akleman2026MagicCubeWebsite,
  author       = {Akleman, Ergun},
  title        = {{Magic Photo Cube}: Interactive Geometric and Textured Simulators},
  year         = {2026},
  howpublished = {Project website, Texas A\&M University, \url{https://people.tamu.edu/~ergun/software/magic_cube/}},
  note         = {Includes interactive geometric and textured simulations and a research preprint. Accessed November 24, 2026}
}

\end{document}